# NEW THEORETICAL RESULTS FOR RADIATIVE $^3$H(p, γ)$^4$He, $^3$He($^2$H, γ)$^5$Li, $^4$He($^2$H, γ)$^6$Li, $^4$He($^3$H, γ)$^7$Li, and $^4$He($^3$He, γ)$^7$Be CAPTURES AT ASTROPHYSICAL ENERGIES

Sergey Dubovichenko[*‡], Albert Dzhazairov-Kakhramanov[*,†]
and Nataliya Burkova[‡]

[*]*Fesenkov Astrophysical Institute "NCSRT" ASA MDASI Republic of Kazakhstan (RK)
Observatory 23, Kamenskoe plato, 050020, Almaty, RK*
[‡]*Al-Farabi Kazakh National University, 050040, av. Al-Farabi 71, Almaty, RK*
[*]*dubovichenko@mail.ru*
[†]*albert-j@yandex.ru*
[‡]*natali.burkova@gmail.com*

**Abstract:** We have studied the proton-capture reaction $^3$H(p,γ)$^4$He. It plays a role in the nucleosynthesis of primordial elements in the early Universe leading to the pre-stellar formation of $^4$He nuclei. All results of our researches and more new data from works show that the contribution of the $^3$H(p,γ)$^4$He capture reaction into the processes of primordial nucleosynthesis is relatively small. However, it makes sense to consider this process for making the picture complete of the formation of prestellar $^4$He and clearing of mechanisms of this reaction. Furthermore, we have considered the $^3$He($^2$H,γ)$^5$Li reaction in the low energy range in the modified potential cluster model with splitting of orbital states according to Young tableaux and, in some cases, with forbidden states. These reaction also form part of the nucleosynthesis chain of the processes occurring in the early stages of formation of stable stars, and are possible candidates for overcoming the well-known problem of the $A = 5$ gap in the synthesis of light elements in the primordial Universe. Continuing study we have considered the radiative capture $^4$He($^3$He, γ)$^7$Be at superlow energies, which has a undeniable interest for nuclear astrophysics, since it takes part in the proton-proton fusion chain, and new experimental data on the astrophysical $S$-factors of this process at energies down to 90 keV and 23 keV and data on the radiative capture reaction $^4$He($^3$H,γ)$^7$Li down to 50 keV appeared recently. Moreover, radiative capture reactions $^4$He($^3$He,γ)$^7$Be and $^4$He($^2$H,γ)$^6$Li may have played a certain role in prestellar nucleosynthesis after the Big Bang, when the temperature of the Universe decreased to the value of 0.3 $T_9$.

**Keywords:** Nuclear astrophysics; primordial nucleosynthesis; light atomic nuclei; low and astrophysical energies; radiative capture; total cross section; thermonuclear processes; potential cluster model; forbidden states.

**PACS Number(s):** 21.60.Gx, 25.20.Lj, 25.40.Lw, 26.20.Np, 26.35.+c, 26.50.+x, 26.90.+n

## 1. Introduction

This review is the logical continuation of our works devoted to the radiative captures on light nuclei that were published in Refs. 1–5.

### 1.1. *Astrophysical aspects of the review*

As we know, light radioactive nuclei play an important role in many astrophysical environments. In addition, such parameter as cross section of the capture reactions as a function of energy and reaction rates are very important for investigation of many astrophysical problems such as primordial nucleosynthesis of the Universe, main trends of stellar evolution, novae and super-novae explosions, X-ray bursts etc. The continued interest in the study of processes of radiative neutron capture on light nuclei at thermal (> 10 meV) and astrophysical (> 1 keV) energies is caused by several reasons. Firstly, this process plays a significant part in the study of many fundamental properties of nuclear reactions, and secondly, the data on the capture cross sections are widely used in a various applications of nuclear physics and nuclear astrophysics, for example, in the process of studying of the primordial nucleosynthesis reactions.

The proton capture on $^3$H reaction is of interest from both theoretical and experimental points of view for understanding the dynamics of photonuclear processes involving the lightest atomic nuclei at low and ultralow, i.e., astrophysical energies.[6] It also plays a role in the nucleosynthesis of primordial elements in the early Universe[6–8] leading to the pre-stellar formation of $^4$He nuclei. Therefore, experimental studies of this reaction continue. New data for the total cross section of proton radiative capture on $^3$H and the astrophysical $S$-factor in the energy range from 50 keV to 5 MeV Ref. 9 and at 12 and 39 keV Ref. 10 in the center of mass system (c.m.) have been obtained. These data will be used by us for further comparison with the calculation results. In addition, we ought to note other experimental studies of the photodisintegration of $^4$He carried out, for example, in Ref. 11. Also, interesting theoretical results for photodisintegration of this nucleus into the p$^3$H channel were published in Ref. 12, including *ab initio* studies (see, for example, Ref. 13).

Upon cooling to a temperature of ~0.8 MeV, the processes of the primordial nucleosynthesis became possible[14,15] with the formation of stable $^2$H, $^3$He and $^4$He nuclei and, also stable in the first minutes of the Universe, $^3$H nucleus. These reactions are shown in Table 1 – the processes of the radiative capture are marked by italic. In table also the data of the $S$-factors and total cross sections at low energies in the energy range 10 – 20 keV were given with references to original works with these results. Table 1 shows that only one of these reactions, No.4, results in energy absorption 0<Q. All of the others lead to energy release Q>0. Some inverse nuclear reactions, for example, photodisintegration of $^{3,4}$He and $^{2,3}$H by gamma-quantum cannot occur because of their extremely low energies at which weak processes cannot keep the balance.[15] Therefore the constant synthesis of stable nuclei without their further disintegration to lighter nuclei becomes possible.

This was the situation when the Universe was about 100 sec old and the number of protons and neutrons was comparable – approximately 0.2 neutrons to each proton. The epoch of primordial nucleosynthesis finished at approximately 200 sec[14] by which time practically all neutrons are bound into $^4$He nuclei and the number of $^4$He is about 25% of the number of $^1$H nuclei. At that point the content of $^2$H and $^3$He relative to $^1$H was about $10^{-4}$–$10^{-6}$.[6–8,15,16]

Thus $^4$He was the last nucleus to emerge at the initial stage of nucleosynthesis because heavier nuclei such as C and O could only be synthesized in the process of nuclear reactions in stars. The reason for this is the existence of some an instability gap for light nuclei ($A = 5$), which, apparently, cannot be bridged in the process of initial nucleosynthesis. In principle, $^4$He could have given rise to heavier nuclei ($A = 7$) in the $^4$He + $^3$H → $^7$Li + γ and $^4$He + $^3$He → $^7$Be + γ reactions. However the Coulomb barrier for these reactions is about 1 MeV while the kinetic energy of the nuclei at temperatures of ~ 1 $T_9$ is of the order of 0.1 MeV and probability of such reactions will be negligible.[17] The mechanism of synthesis of $^4$He explains its abundance in the Universe confirms its origin at the pre-stellar stage and corroborates the Big Bang theory.

It is important to estimate the $S$-factors of reactions 1–14. For example, as will be seen further, the astrophysical $S$-factor of proton capture on $^2$H at an energy of 1 keV is in 5–10 times lower than the $S$-factor of the proton capture on $^3$H at the same energy.[20] This means that the latter process, which contributes to the formation of $^4$He in primordial nucleosynthesis, is much more likely, in spite of the lower abundance of $^3$H relative to $^2$H.[14,15,32] Most data available in the literature[6–9,17,20] relate to the abundance of elements such as $^3$He at present time. This is generally confirmed by modern astrophysical observations.[14,32] However, the abundance of $^3$H for the first 100–200 s after the Big Bang cannot be much smaller than that of $^2$H since the neutron capture reaction, in spite of the reduction of neutron numbers down to 0.2 of the proton numbers, can go on deuteron at any energy. In addition, the half-life of $^3$H is 4500(8) days[33] and do not make a real contribution to the decrease of the number of $^3$H at the first few minutes after the Big Bang.

Table 1. Basic reaction of the primordial nucleosynthesis with light nuclei.[16]

| No. | Process | Released energy in MeV | Astrophysical $S$-factor in keV b at 10 – 20 keV in center of mass – the accurate energy is stated in square brackets | The total cross section $\sigma_t$ in µb for the given energy in square brackets | Reference |
|---|---|---|---|---|---|
| 1. | $p+n \rightarrow {}^2H+\gamma$ | 2.225 | 3.18(25)·10$^{-3}$ [10.0] | 3.18(25)·10$^2$ [10.0] | Ref. 18 |
| 2. | $^2H+p \rightarrow {}^3He+\gamma$ | 5.494 | 3.0(6)·10$^{-4}$ [10.4] | 1.0(2)·10$^{-2}$ [10.4] | Ref. 19 |
| 3. | $^2H+n \rightarrow {}^3H+\gamma$ | 6.257 | 1.2·10$^{-5}$ [10.5]* | 1.1 [10.5]* | Ref. 20 |
| 4. | $^3H+p \rightarrow {}^3He+n$ | –0.763 (see Ref. 21) | 2536 [12]*** | 81537 [roughly at 12 keV above the threshold or 1.03354 MeV in l.s.] | Ref. 22 |
| 5. | $^3He+n \rightarrow {}^3H+p$ | 0.764 | 63.2 [10.3] | 6.14(16) ·10$^6$ [10.3] | Ref. 23 |
| 6. | $^3H+p \rightarrow {}^4He+\gamma$ | 19.814 | 2.2·10$^{-3}$ [10.0] | 4.0·10$^{-2}$ [10.0] | Ref. 10 |
| 7. | $^3He+n \rightarrow {}^4He+\gamma$ | 20.578 | 1.7·10$^{-4}$ [18.4] | 9.2(2.0) [18.4] | Ref. 24 |
| 8. | $^2H+{}^2H \rightarrow {}^3He+n$ | 3.269 | 51.4(2.0) [9.94]<br>53.05(0.55) [10.0]*** | 241.3(9.4) [9.94]**<br>255.1(2.9) [10.0] | Ref. 25<br>Ref. 26 |
| 9. | $^2H+{}^2H \rightarrow {}^3H+p$ | 4.033 | 56.1(1.6) [9.97] | 270.4(7.6) [9.97] | Ref. 27 |
| 10. | $^2H+{}^3He \rightarrow {}^4He+p$ | 18.353 | 7480(200) [10.7] | 0.5(1) [10.7]** | Ref. 28 |
| 11. | $^2H+{}^3H \rightarrow {}^4He+n$ | 17.589 | 12328.4 [9]*** | 14200 [9] | Ref. 29 |
| 12. | $^2H+{}^2H \rightarrow {}^4He+\gamma$ | 23.847 | 5.7(2.4)·10$^{-6}$ [10.0] | 2.9(1.2)·10$^{-5}$ [10.0] | Ref. 30 |
| 13. | $^2H+{}^3He \rightarrow {}^5Li+\gamma$ | 16.66 | 0.41 [111]*** | 5.3 [111] | Ref. 31 |
| 14. | $^2H+{}^3H \rightarrow {}^5He+\gamma$ | 16.792 | 0.17 [90]*** | 50 [90] | Ref. 31 |

\* - theoretical value calculated on the basis of the Modified Potential Cluster Model
\*\* - the value calculated on the basis of the $S$-factor
\*\*\* - the value calculated on the basis of the total cross section

The quantity of tritium, additionally to process No.3, also increases due to reactions No.5 and No.9, but can decrease due to processes No.6 and 11. At energies lower than 0.8 MeV reaction No.4 makes virtually no contribution to reductions in tritium. Meanwhile, the total cross section of reaction No.11 is about 14.2 mb at 9 keV Ref. 29 and of the reaction No.6 is about 4·10$^{-2}$ µb at 10 keV Ref. 10 show their small relative contributions to the formation of $^4$He. However the number of deuterons available for reaction No.11 is approximately 4–5 orders of magnitude less than the number of protons taking part in reaction No.6. Therefore the overall contribution of the two reactions in pre-stellar formation of $^4$He will be similar.

Reaction No.12 proceeds with comparatively low probability, since the $E$1 process is forbidden by the isospin selection rules. This leads to the factor $(Z_1/m_1^J + (-1)^J Z_2/m_2^J)$ at multipolarity of γ-quantum of $J = 1$.[20] This product defines the value of the total cross sections of the radiative capture and $E$1 processes with the same Z/m ratio, for particles of the initial channel leads to zero cross sections. The probability of the allowed $E$2 transitions in such processes is usually nearly 1.5 to 2.0 orders of magnitude less[34] that was noted earlier in reviews.[14,15] This fact well demonstrates cross section's data for reaction No.12 from Table 1, the value of which is lowest. For two last reactions No.13, 14 we cannot find experimental data at lower energies.

Let us show furthermore the reaction rates given in Ref. 35 in the form of parametrizations. Shape of these rates for first reactions, leading to the formation of $^4$He or nuclei with mass of 3, is shown in Fig. 1. One can see that considered reaction is at the third level for rate of forming $^4$He and its rate in some times lower, for example, that the reaction rates of $^3$H(d,n)$^4$He or $^3$He(d,p)$^4$He. However, at the energy about 10 $T_6$ the rate of the last reaction equals the reaction rate of the proton radiative capture on $^3$H.

All these results and more new data from Refs. 36,37 show that the contribution of the $^3$H(p,γ)$^4$He capture reaction into the processes of primordial nucleosynthesis is relatively small. However, it makes sense to consider this process for making the picture complete of the formation of prestellar $^4$He and clearing of mechanisms of this reaction. In addition, as it was shown furthermore, our calculations of this reaction rate, based on the modern data of the astrophysical $S$-factors,[10] lays slightly lower from the results of works.[35,38–40] The latest works do not take into account new data,[9,10] which were taken into account by us in this work, and our results can be considered as an improved data on the rate of the considered reaction.

Moreover it should be noted that our understanding of the different stages in the formation of the Universe, of the processes of nucleosynthesis occurring in it and of the properties of new stars, is still developing. Therefore there is a pressing need to acquire new information on primordial nucleosynthesis and on the mechanisms of the Universe's formation and this is one of the main tasks for the construction of a unified cosmological model. All of this directly applies to the detailed study of the p$^3$H capture reaction in the astrophysical energy region on the basis of the modern nuclear model. This model, as shown below, has already demonstrated its efficiency in the description of the characteristics of almost 30 such reactions.[4,20,41–45]

Continuing study of thermonuclear reactions in the frame of the modified potential cluster model (MPCM) with forbidden states (FSs) let us consider the radiative capture $^4$He($^3$He, γ)$^7$Be at superlow energies has a undeniable interest for nuclear astrophysics, since it takes part in the proton-proton fusion chain, and new experimental data on the astrophysical $S$-factors of this process at energies down to 90 keV and 23 keV and data on the radiative capture reaction $^4$He($^3$H, γ)$^7$Li down to 50 keV appeared recently. The proton-proton chain may be completed by the following process with a probability of 86%:[46]

$$^3He + {}^3He \rightarrow {}^4He + 2p$$

or the reaction, considered here, involving the prestellar $^4$He, (see, for example, Ref. 8)

$$^3\text{He} + {}^4\text{He} \rightarrow {}^7\text{Be} + \gamma, \qquad (1)$$

the probability of which is 14%.[46] Moreover, radiative capture reactions $^4\text{He}(^3\text{He},\gamma)^7\text{Be}$ and $^4\text{He}(^2\text{H},\gamma)^6\text{Li}$ may have played a certain role in prestellar nucleosynthesis after the Big Bang, when the temperature of the Universe decreased to the value of 0.3 $T_9$ Ref. 47 ($T_9 = 10^9$ K).

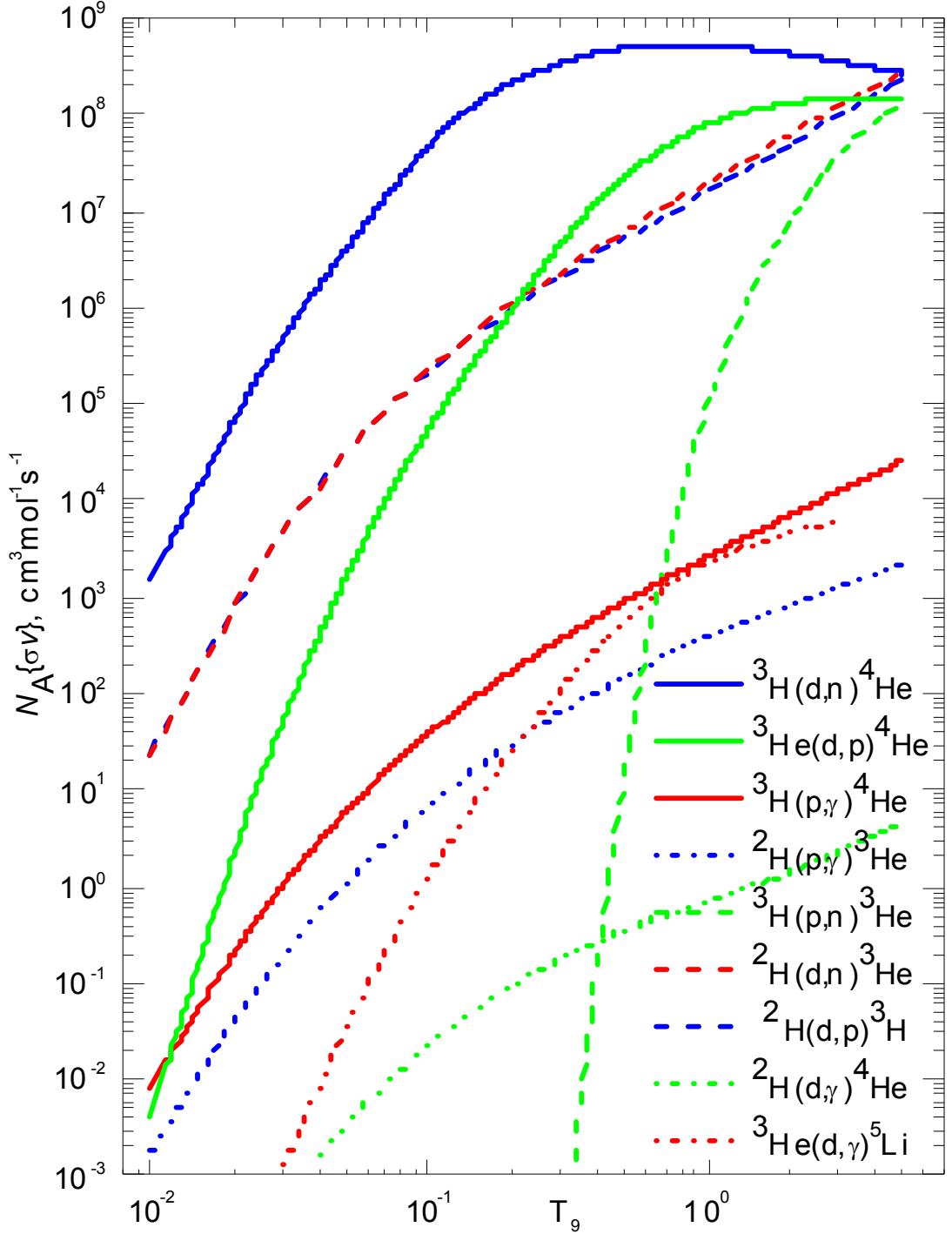

Fig. 1. Reaction rates from Refs. 16,35.

## 1.2. *Nuclear aspects of the review*

One extremely successful line of development of nuclear physics in the last 50-60 years has been the microscopic model known as the Resonating Group Method (RGM, see, for example, Refs. 48–52). And the associated with it models, for example, Generator Coordinate Method (see, particularly, Refs. 52,53) or algebraic version of RGM Refs. 54,55. However, the rather difficult RGM calculations are not the only way in which to explain the available experimental facts. But, the possibilities offered by a simple two-body potential cluster model (PCM) have not been studied fully up to now, particularly if it uses the concept of FSs.[56] The potentials of this model for discrete spectrum are constructed in order to correctly reproduce the main characteristics of the BSs of light nuclei in cluster channels, and in the continuous spectrum they directly take into account the resonance behavior of the elastic scattering phase shifts of the interactive particles at low energies.[57,58] It is enough to use the simple PCM with FSs taking into account the described methods of construction of potentials and classification of the orbital states according to Young tableaux for consideration many problems of nuclear physics of low energy and nuclear astrophysics. Such a model can be called a modified PCM. In many cases, such an approach, as has been shown previously, allows one to obtain adequate results in the description of many experimental studies for the total cross sections of the thermonuclear reactions at low and astrophysical energies.[56–58]

Therefore, in continuing to study the processes of radiative capture,[57,58] we will consider the p+$^3$H→$^4$He+γ, $^2$H$^3$He → $^5$Liγ, $^2$H$^4$He → $^6$Liγ, $^3$H$^4$He → $^7$Liγ, $^3$He$^4$He → $^7$Beγ reactions within the framework of the MPCM at astrophysical and low energies. The resonance behavior of the elastic scattering phase shifts of the interacting particles at low energies will be taken into account. In addition, the classification of the orbital states of the clusters according to the Young tableaux allows one to clarify the number of FSs and allowed states (ASs), i.e., the number of nodes of the wave function (WF) of the relative motion of the cluster.

Furthermore, let us consider the $^3$He($^2$H,γ)$^5$Li reaction in the low energy range in the modified potential cluster model (MPCM) with splitting of orbital states according to Young tableaux and, in some cases, with FSs. Our interest in the radiative capture reactions in the isobar-analogue channels $^3$H($^2$H,γ)$^5$He and $^3$He($^2$H,γ)$^5$Li is due to following two main reasons. New data may be found in Ref. 59 in relation to the diagnostics of the nuclear fusion efficiencies of $^3$H($^2$H,n)$^4$He and $^3$He($^2$H,p)$^4$He reactions, used for study of Tokamak plasmas in experiments on JET and ITER. These reactions also form part of the nucleosynthesis chain of the processes occurring in the early stages of formation of stable stars, and are possible candidates for overcoming the well-known problem of the $A = 5$ gap in the synthesis of light elements in the primordial Universe.[6]

The present paper reports on the treatment of the $^3$He($^2$H,γ)$^5$Li reaction, which has been insufficiently studied. Nearly all of the scarce theoretical models of this channel are based on data from a single experimental study by Buss, carried out in 1968 Ref. 31 on the total cross section of the deuteron radiative capture on $^3$He in the deuteron energy range 200–1400 keV. The most complete nuclear database is EXFOR,[60] and the known open-access databases of nuclear characteristics, PHYSICS, CDFE, NASA DATA (see, for example, Refs. 61,62) contain only the same data. There are also no consistent and detailed studies of the $^3$He($^2$H,γ)$^5$Li reaction which include the dynamic and statistical characteristics of the continuous and bound states of the $^2$H$^3$He cluster system.

However, we have successfully used the MPCM based on the inclusion of the Pauli forbidden states with a corresponding classification according to Young tableaux.[5,20,41,43,44,63] This model is much more simple and transparent than the microscopic resonating group method (RGM).[64,65] MPCM enables consistent numerical results to be obtained for the majority of the total cross-sections of the radiative capture reactions and rates at astrophysical and thermonuclear energies for the more than 30 processes treated in Refs. 5,20,41,43,44,57,63, as well as binding energies and low-lying excited spectra, root mean square charge and mass radii, and asymptotic normalizing coefficients (ANC) in cluster channels.

Here we present new calculated data for the root mean square radii and ANC for $^5$Li in the $^2$H$^3$He channel. The total cross section of the deuteron radiative capture on $^3$He to the ground state (GS) of $^5$Li for the dipole $E$1 and $M$1 transitions both in doublet and quartet spin channels have also been calculated. For the obtained astrophysical $S$-factor and reaction rate the analytical parametrizations were found as functions of $E$ and $T_9$ respectively.

## 2. Model and calculation methods

The nuclear part of the intercluster interaction potential, which depends on set of quantum numbers $JLS$, for carrying out calculations of photonuclear processes in the considered cluster systems, has the form:

$$V_{JLS\{f\}}(R) = V_0(JLS\{f\})\exp[-\alpha(JLS\{f\})R^2] + V_1(JLS\{f\})\exp[-\gamma(JLS\{f\})R], \qquad (2)$$

$$V_{Coul}(r) = \begin{cases} \dfrac{Z_1 Z_2}{r} & r > R_{Coul} \\ Z_1 Z_2 \left(3 - \dfrac{r^2}{R_{Coul}^2}\right) \Big/ 2R_{Coul} & r < R_{Coul} \end{cases}.$$

with Coulomb term of the potential of spherical or point-like form.

The potential is constructed completely unambiguously with the given number of BSs and with the analysis of the resonance scattering when in the considered partial wave at energies up to 1 MeV where there is a rather narrow resonance with a width of about 10–50 keV. Its depth is unambiguously fixed according to the resonance energy of the level at the given number of BS, and the width is absolutely determined by the width of such resonance. The error of its parameters does not usually exceed the error of the width determination at this level and equals 3–5%. Furthermore, it concerns the construction of the partial potential according to the phase shifts and determination of its parameters according to the resonance in the nuclear spectrum.

Consequently, all potentials do not have ambiguities and allow correct description of total cross sections of the radiative capture processes, without involvement of the additional quantity – spectroscopic factor $S_f$.[66] It is not required to introduce additional factor $S_f$ under consideration of capture reaction in the frame of PCM for potentials that are matched, in continuous spectrum, with characteristics of scattering processes that take into account resonance shape of phase shifts, and in the discrete spectrum, describing the basic characteristics of nucleus BS.

All effects that are present in the reaction, usually expressed in certain factors and coefficients, are taken into account at the construction of the interaction potentials. It could be possible, exactly because they are constructed and take into account FS structure. On the basis of description of observed, i.e., experimental characteristics of interacting clusters in the initial channel and formed, in the final state, a certain nucleus that has a cluster structure consisting of initial particles. In other words, the presence of $S_f$, is apparently taken into account in the BS WFs of clusters, determining the basis of such potentials due to solving the Schrödinger equation.[67]

The AC for any GS potential was calculated using the asymptotics of the WF having a form of exact Whittaker function[68]

$$\chi_L(r) = \sqrt{2k_0} C_W W_{-\eta L+1/2}(2k_0 r), \quad (3)$$

where $\chi_L(r)$ is the numerical WF of the BS, obtained from the solution of the radial Schrödinger equation and normalized to unity, the value $W_{-\eta L+1/2}$ is the Whittaker function of the BS, determining the asymptotic behavior of the WF, which is the solution of the same equation without the nuclear potential. $k_0$ is the wave number, caused by the channel binding energy $E$: $k_0 = \sqrt{2\mu \frac{m_0}{\hbar^2} E}$; $\eta$ is the Coulomb parameter $\eta = \frac{\mu Z_1 Z_2 e^2}{\hbar^2 k}$, determined numerically $\eta = 3.44476 \cdot 10^{-2} \frac{\mu Z_1 Z_2}{k}$ and $L$ is the orbital angular momentum of the BS. Here $\mu$ is the reduced mass, and the constant $\hbar^2/m_0$ is assumed to be 41.4686 fm$^2$, where $m_0$ is the atomic mass unit (amu). The magnetic moment of neutron equals $\mu = -1.91304272\mu_0$, and $1.653560\mu_0$ for $^8$Li,[69] where $\mu_0$ is the nuclear magneton. Slightly transformed expressions[20] of Refs. 70,71 were used for total cross sections of the electromagnetic transitions.

Data on the spectroscopic factor $S$ of the GS and the asymptotic normalization coefficients $A_{NC}$ (ANC) are given, for example, in Ref. 72. Here we also use the relationship

$$A_{NC}^2 = S \times C^2, \quad (4)$$

where $C$ is the asymptotic constant in fm$^{-1/2}$, which is related to the dimensionless AC $C_W$,[68] used by us in the following way: $C = \sqrt{2k_0} C_W$.

The total radiative capture cross sections $\sigma(NJ, J_f)$ for the $EJ$ and $MJ$ transitions in the case of the PCM are given, for example, in Ref. 66 or Refs. 57,58,73,74 are written as:

$$\sigma_c(NJ, J_f) = \frac{8\pi K e^2}{\hbar^2 q^3} \frac{\mu}{(2S_1+1)(2S_2+1)} \frac{J+1}{J[(2J+1)!!]^2}$$
$$\times A_J^2(NJ, K) \sum_{L_i, J_i} P_J^2(NJ, J_f, J_i) I_J^2(J_f, J_i) \quad (5)$$

where $\sigma$ – total radiative capture cross section; $\mu$ – reduced mass of initial channel particles; $q$ – wave number in initial channel; $S_1$, $S_2$ – spins of particles in initial channel; $K, J$ – wave number and momentum of $\gamma$-quantum in final channel; $N$ – is the $E$ or $M$ transitions of the $J$ multipole ordered from the initial $J_i$ to the final $J_f$ nucleus state.

The value $P_J$ for electric orbital $EJ(L)$ transitions has the form Refs. 57,58

$$P_J^2(EJ, J_f, J_i) = \delta_{S_i S_f} [(2J+1)(2L_i+1)(2J_i+1)(2J_f+1)](L_i 0 J 0 | L_f 0)^2 \begin{Bmatrix} L_i & S & J_i \\ J_f & J & L_f \end{Bmatrix}^2,$$

$$A_J(EJ,K) = K^J \mu^J \left( \frac{Z_1}{m_1^J} + (-1)^J \frac{Z_2}{m_2^J} \right), \quad I_J(J_f, J_i) = \langle \chi_f | r^J | \chi_i \rangle. \quad (6)$$

Here, $S_i$, $S_f$, $L_f$, $L_i$, $J_f$, and $J_i$ – total spins, angular and total moments in initial (*i*) and final (*f*) channels; $m_1$, $m_2$, $Z_1$, $Z_2$ – masses and charges of the particles in initial channel; $I_J$ – integral over WFs of initial $\chi_i$ and final $\chi_f$ states, as functions of cluster relative motion of n and $^{10}$B particles with intercluster distance $R$.

For consideration of the $M1(S)$ magnetic transition, caused by the spin part of magnetic operator,[34] it is possible to obtain an expression[57,58] using the following Ref. 75:

$$P_1^2(M1, J_f, J_i) = \delta_{S_i S_f} \delta_{L_i L_f} \left[ S(S+1)(2S+1)(2J_i+1)(2J_f+1) \right] \begin{Bmatrix} S & L & J_i \\ J_f & 1 & S \end{Bmatrix}^2,$$

$$A_1(M1, K) = i \frac{\hbar K}{m_0 c} \sqrt{3} \left[ \mu_1 \frac{m_2}{m} - \mu_2 \frac{m_1}{m} \right], \quad I_J(J_f, J_i) = \langle \chi_f | r^{J-1} | \chi_i \rangle \quad J=1. \quad (7)$$

Here, $m$ is the mass of the nucleus, and $\mu_1$ and $\mu_2$ are the magnetic moments of the clusters, the values of which are taken from Refs. 76,77.

The construction methods used here for intercluster partial potentials at the given orbital moment $L$, are expanded in Refs. 57,58,78 and here we will not discuss them further. The next values of particle masses are used in the given calculations: $m_p = 1.00727646577$ amu,[79] $m(^2\text{H}) = 2.014102$ amu,[80] $m(^3\text{H}) = 3.016049$ amu,[81] $m(^3\text{He}) = 3.016029$ amu,[82] $m(^4\text{He}) = 4.002603$ amu,[83] $m(^6\text{Li}) = 6.015123$ amu,[84] $m(^7\text{Li}) = 7.016005$ amu,[85] $m(^7\text{Be}) = 7.016930$ amu,[86] and constant $\hbar^2/m_0$ is equal to 41.4686 MeV fm$^2$.

## 3. Capture reaction $^3$H(p, γ)$^4$He

### 3.1. Classification of p$^3$H states according to Young tableaux

The preliminary studying of the radiative capture reaction $^3$H(p, γ)$^4$He was published by us in Ref. 16. It is known Ref. 87 that states with minimal spin in scattering processes in the certain lightest atomic nuclei are mixed with respect to orbital Young tableaux, for example, the singlet state of p$^3$H system is mixed according to tableaux {4} and {31}.[16,88] At the same time, this state in the bound form, for example, singlet p$^3$H channel of $^4$He is the pure state with Young tableau {4}.[88] In this case we can suppose[87] that BSs and scattering potentials for $N^3$H ($N^3$He) states will be different because of the difference of their Young tableaux. Thus, the explicit dependence of the potential parameters at the given moments $L$, $S$ and $J$ from Young tableaux {*f*} is permitted in this case.

Now, let's give the classification of states, for example, of $N^3$H ($N^3$He) systems according to orbital and spin-isospin Young tableaux and demonstrate how to obtain these results. In the general case, the possible orbital Young tableau {*f*} of some nucleus $A(\{f\})$ consisting of two parts $A_1(\{f_1\}) + A_2(\{f_2\})$ is the direct outer product of orbital Young tableaux of these parts $\{f\}_L = \{f_1\}_L \times \{f_2\}_L$ and is determined by the Littlewood theorem.[87,88] Therefore, the possible orbital Young tableaux of the $N^3$H ($N^3$He) systems, in which tableau {3} is used for $^3$H ($^3$He), are the symmetries $\{4\}_L$ and $\{31\}_L$.

Spin-isospin tableaux are the direct inner product of spin and isospin Young tableaux of the nucleus of $A$ nucleons $\{f\}_{ST} = \{f\}_S \otimes \{f\}_T$ and for the system with the number of particles not larger than eight are given in Ref. 89. For any of these moments (spin and isospin), the corresponding tableau of the nucleus consisting of $A$ nucleons each of which has an angular moment equals 1/2 is constructed as follows: in the cells of the first row the number of nucleons with the moments pointing in one direction, for example, upward, is indicated. In cells of the second row, if it is required, the number of nucleons with the moments directed in the opposite direction, for example, downward, is indicated. The total number of cells in both rows is equal to the number of nucleons in the nucleus. Moments of nucleons in the first row which have a pair in the second row with the oppositely directed moment are compensated and have, therefore, a zero total moment. The sum of moments of nucleons of the first row, which are not compensated by moments of nucleons of the second one, gives the total moment of the whole system.[90]

In this case for $N\,^3$H ($N\,^3$He) cluster systems at the isospin $T = 0$ and the spin $S = 0$, we have tableau $\{22\}_S$ or $\{22\}_T$; and for $S$ or $T = 1$, the Young tableau has the form $\{31\}_S$ or $\{31\}_T$. Upon construction of the spin-isospin Young tableau for the triplet spin state of $N\,^3$H ($N\,^3$He) systems with $T = 1$, we have $\{31\}_S \otimes \{31\}_T = \{4\}_{ST} + \{31\}_{ST} + \{22\}_{ST} + \{211\}_{ST}$, and for the singlet spin state with $T = 0$, we have $\{22\}_S \otimes \{22\}_T = \{4\}_{ST} + \{22\}_{ST} + \{1111\}_{ST}$.[89]

The total Young tableau of the nucleus is determined in a similar way as the direct inner product of the orbital and spin-isospin tableau $\{f\} = \{f\}_L \otimes \{f\}_{ST}$.[87] The total wave function of the system in the case of antisymmetrization does not identically vanish only if it does not contain the antisymmetric component $\{1^N\}$, that is realized upon multiplication of conjugated $\{f\}_L$ and $\{f\}_{ST}$. Therefore, the tableaux $\{f\}_L$ conjugated to $\{f\}_{ST}$ are allowed in this channel and all other symmetries are forbidden, since they result to zero total wave function of the system of particles after its antisymmetrization.

Thus, for p$^3$H system in the triplet channel, independently from the $T$ values, only the orbital wave function with the symmetry $\{31\}_L$ is allowed and the function with $\{4\}_L$ turns out to be forbidden, since the products $\{211\}_{ST} \otimes \{4\}_L$ or $\{31\}_{ST} \otimes \{4\}_L$ do not result in an antisymmetric component of the total wave function. At the same time, in the singlet channel for $T = 0$, we have $\{1111\}_{ST} \otimes \{4\}_L = \{1111\}$,[89] and we obtain the antisymmetric tableau. At this channel with $T = 1$ we have the product $\{211\}_{ST} \otimes \{31\}_L$ that also gives the antisymmetric component $\{1111\}$ for total wave function. Just that very case when one can conclude that the singlet spin state for p$^3$H and n$^3$He systems turns out mixed according to Young orbital tableaux each of which relates to different isospin values.

In other words, p$^3$H system is mixed with respect to isospin, since it has the projection $T_z = 0$, and the following values of the total isospin are possible: $T = 0$ and 1. Hence, in this system both triplet and singlet phase shifts and, therefore, potentials effectively depend on two isospin values. Mixing with respect to isospin leads to mixing according to Young tableaux. As it was shown above, in the singlet spin state two orbital Young tableaux – $\{31\}$ and $\{4\}$ – are allowed.[91] Then it was shown in Ref. 16,88,91 that singlet phase shifts of the p$^3$H scattering mixed with respect to isospin can be represented in the form of the half-sum of pure with respect to isospin singlet phase shifts

$$\delta^{\{T=1\}+\{T=0\}} = 1/2[\delta^{\{T=1\}} + \delta^{\{T=0\}}], \tag{8}$$

this is equivalent to the following expression for the scattering phase shifts in terms of Young tableaux

$$\delta^{\{4\}+\{31\}} = 1/2[\delta^{\{31\}} + \delta^{\{4\}}]. \tag{9}$$

Pure phase shifts with Young tableau {31} correspond to $T = 1$, and phase shifts with {4} to isospin $T = 0$. In this approach we assume that pure phase shifts with isospin $T = 1$ in p$^3$H system can be matched to phase shifts with $T = 1$ for p$^3$He channel.[88,91] Therefore p$^3$He system at $T_z = 1$ is pure by isospin with $T = 1$, so the pure by isospin phase shifts of p$^3$H scattering with $T = 0$ are extracted from expression (8) on the basis of the known pure scattering phase shifts with $T = 1$ for p$^3$He system[92–97] and for mixed p$^3$H with isospin with $T = 0$ and 1.[98–100] Furthermore, the corresponded pure potentials of p$^3$H interaction are constructed on their basis, for example, for the GS of $^4$He in p$^3$H channel.[88,91]

## 3.2 Potentials for p$^3$H and p$^3$He systems

For calculations of the photonuclear processes in the system considered the nuclear part of the intercluster potential of p$^3$He interactions for each partial wave can be expressed in form (2) with a point-like Coulomb term. This potential, as for that of the p$^2$H system,[20,42] is constructed so as to correctly describe the corresponding partial phase shift of the p$^3$He elastic scattering.[91,101]

Consequently the pure (with respect to isospin $T = 1$) potentials of p$^3$He interactions for the elastic scattering processes were obtained, and their parameters are listed in (10) and (11) Refs. 41,88. The singlet potentials of the form given in expression (2) for the p$^3$He scattering, pure with respect to isospin $T = 1$:[41,88]

$$p^3He\ System$$
$$^1S\ \text{wave} - V_0 = -110.0\ \text{MeV},\ \alpha = 0.37\ \text{fm}^{-2},\ V_1 = +45.0\ \text{MeV},\ \gamma = 0.67\ \text{fm}^{-1}, \tag{10}$$
$$^1P\ \text{wave} - V_0 = -15.0\ \text{MeV},\ \alpha = 0.1\ \text{fm}^{-2}. \tag{11}$$

Note that this singlet and pure (with respect to isospin) $S$ phase shift of the p$^3$He elastic scattering is used further for calculation of the singlet p$^3$H phase shifts with the isospin $T = 0$. The singlet $^1P_1$ phase shift of the p$^3$He elastic scattering with $T = 1$ used in our calculations of the $E1$ transition to the ground state of $^4$He in the p$^3$H channel with $T = 0$. The scattering phase shifts obtained with such potentials are given in our previous Refs. 91,101.

The singlet, isospin and Young tableaux mixed $S$ phase shift of the elastic p$^3$H scattering, determined from the experimental differential cross sections, and used later for obtaining the pure p$^3$H phase shifts for potential (2) at $V_1 = 0$ with parameters

$$V_0 = -50\ \text{MeV},\ \alpha = 0.2\ \text{fm}^{-2}. \tag{12}$$

Then, the following parameters at $V_1 = 0$ for the pure p$^3$H potential with $T = 0$ in the $^1S$ wave in Refs. 91,101 have been found:

$$V_0 = -63.1 \text{ MeV}, \quad \alpha = 0.17 \text{ fm}^{-2}. \tag{13}$$

Scattering phase shifts for potentials (12) and (13) are given in Refs. 16,91,101. The pure (according to Young tableaux) interactions thus obtained can be used for the calculation of different characteristics of the bound ground state $^4$He in the p$^3$H channel. The degree of agreement of the results obtained in this case with experiment now depends only on the degree of clusterization of this nucleus in the channel considered and here one supposes that this degree is high enough that the spectroscopic factor of the channel will be close to unity.

The interaction potential (13) obtained in Refs. 91,101 on the whole correctly describes the channel binding energy of the p$^3$H system (to several keV) and the root-mean-square radius of $^4$He. Using this potential and the potential of the $^1P$ scattering wave with the point-like coulomb term for the p$^3$H system from (11), the differential and total cross sections of proton radiative capture on $^3$H were calculated in Refs. 91,101 and Ref. 88 respectively. The astrophysical $S$-factors at energies down to 10 keV were also calculated.

It should be noted that at that time experimental data for the $S$-factor only was known in the energy region above 700–800 keV.[102] Subsequently new experimental data were obtained in Ref. 9 and Ref. 10. It will be of interest to explore whether the potential cluster model with the singlet $^1P$ potential obtained earlier and the refined interaction of the pure ground $^1S$ state of $^4$He is capable of describing this new more accurate data.

Our preliminary results Refs. 88,101 have shown that for calculation of the $S$-factor at energies of the order of 1 keV the same conditions as in the p$^2$H system[20,42] should be satisfied. In particular the accuracy of values obtained for the binding energy of $^4$He in the p$^3$H channel should be increased. New modified programs as described in Ref. 41,42 were used in the current work in order to improve parameters of the potential of the ground state for the p$^3$H system of $^4$He as given in Ref. 103.

The results for pure potentials (with respect to isospin of $T = 0$) are given in the expression (2) with the following parameters:

$$p^3H \text{ System}$$
$$^1S \text{ wave} - V_0 = -62.906841138 \text{ MeV}, \quad \alpha = 0.17 \text{ fm}^{-2}. \tag{14}$$
$$^1P \text{ wave} - V_0 = +8.0 \text{ MeV}, \quad \alpha = 0.03 \text{ fm}^{-2}. \tag{15}$$

These results obtained in Ref. 91 differ from those presented in Ref. 101 by approximately 0.2 MeV. This difference is mainly connected with the use, in the new calculations, of more accurate values of masses of p and $^3$H particles[104] and more accurate description of the binding energy of $^4$He in the p$^3$H channel. Using this value of $-19.813810$ MeV was obtained.[21] The calculation with the potential considered here gives $-19.81381000$ MeV.

The behavior of the "tail" of the numerical wave function (WF) $\chi_L(R)$, for the p$^3$H system bound state at large distances, was verified using an asymptotic constant (AC) $C_W$ (3).[68,105] The reduced error of the asymptotic constant $C_w$ was achieved by averaging it over the range in which its variation is a minimum. To find the region for $C_w$ stabilization we calculated it starting at the maximum distances that were considered by us of about 20–30 fm. This stabilization usually occurs at distances of about 7–12 fm. In this region the $C_W$ changes least with a variation of about $10^{-3}$. At distances below the stabilization region for the WF $\chi_L(R)$ we use numerical values of this function obtained from the Schrödinger equation solution. At distances above the stabilization region it is calculated from its asymptotic (3) determined by the Whittaker function $W_{-\eta L+1/2}(2k_0 r)$ using the value for $C_w$ found in the stabilization region.

The experimentally determined value of $C_W$ in Ref. 68 is 5.16(13) for the p$^3$H channel. For the n$^3$He system a value of 5.10(38) was obtained. This is very close to the constant of the p$^3$H channel for our GS potential from (14). Ref. 105 reports a value of 4.1 constant of the n$^3$He system and 4.0 for p$^3$H. The average value between the results of Refs. 105 and 68 are in good agreement with our results for the GS potential from (14). Apparently, there is a considerable difference between the data of asymptotic constants. For the n$^3$He system reported values range from 4.1 to 5.5 and for the p$^3$H channel from 4.0 to 5.3.

The results reported in Ref. 106, where the average spectroscopic $S_f$ factor was 1.59 and the average value of the asymptotic normalizing coefficient $A_{NC}$ (ANC) was 6.02 fm$^{-1/2}$, were obtained on the basis of calculations with different potentials. The relationship between ANC and dimensional AC $C$:[103,107]

$$A_{NC}^2 = S_f \times C^2 \qquad (16)$$

where $C$ can be found from

$$\chi_L(r) = C W_{-\eta L+1/2}(2k_0 r) \qquad (17)$$

This dimensional constant is related to the non-dimensional $C_W$ used by us by $C = \sqrt{2k_0} C_W$. So using the values of ANC and $S_f$ given in Ref. 106 we obtained a value for $C$ of 4.77 fm$^{-1/2}$. In this case $\sqrt{2k_0} = 1.30$ so the dimensionless AC $C_w$ is 3.67. This is slightly less than values obtained here and given in Ref. 68,105. However, if the spectroscopic factor determines the possibility of certain two-body channel, then it is unlikely to be more than unity. $S_f = 1.0$ would give a $C$ of 6.02 and $C_W$ of 4.63. This agrees acceptably with the given above dimensionless value of 4.52 for the $^1S$ potential of the ground state from (14).

For the charge radius of $^4$He, with the potential from (14), we have obtained a value of 1.78 fm (calculation methods for charge radius are described in Refs. 41,42,88,101. The experimentally determined value of $^4$He radius 1.671(14) fm.[21] For these calculations we have used the values of the tritium radius of 1.73 fm from Ref. 33 and the proton radius of 0.8775 fm from data base Ref. 104.

### *3.3 Astrophysical S-factor and reaction rate*

For calculation of the astrophysical *S*-factor the usual expression was used (see, for example, Ref. 38)

$$S = \sigma E \exp(31.335 Z_1 Z_2 \sqrt{\frac{\mu}{E}}), \qquad (18)$$

where σ is the total cross section in barn, $E$ is the center-of-mass energy in keV, $Z_i$ are the particle charges, μ is the reduced mass of particles in amu.[6]

The total cross sections and the astrophysical S-factor of the proton radiative capture process on $^3$H have previously been calculated based on the modified potential cluster model.[101] It was assumed that the main contributions to the cross sections of $E$1 photodisintegration of $^4$He in the p$^3$H channel, or to the proton radiative capture on $^3$H, were transition with changing isospin by the unit for which $\Delta T = 1$.[108] Therefore our calculations will assume that the $^1P_1$ potential for p$^3$He scattering is pure with respect to the isospin ($T = 1$) singlet state of this system and that the $^1S$ potential for the ground state is pure with respect to the isospin $T = 0$ bound state of $^4$He in the p$^3$H channel.[16]

Using these assumptions, the $E$1 transition with refined potential of the ground state of $^4$He, as shown in (14), was re-calculated. The results for the astrophysical $S$-factor at energies from 1 keV up to 5 MeV are shown in Fig. 2 by the green solid line. In particular the new results obtained for the energy region from 10 keV to 5 MeV are in close agreement with our previous results as given in Ref. 91.

Fig. 2 also shows the resulting values from new experimental data Refs. 9,10 and additional data from Ref. 109 not known to us earlier. It can be seen from this figure that the calculations performed about 20 years ago[101] well reproduce the data on the $S$-factor obtained in Ref. 9 at energies of p$^3$H capture from 50 keV to 5 MeV (c.m.). These data were published after the publication of our article Ref. 101 and have noticeably lower ambiguity at energies lower 5 MeV (Fig. 2) than do earlier results Refs. 102,110–112 and they more accurately determine the general behavior of the $S$-factor at low energies, practically coinciding with early data Ref. 109 in the energy range 80–600 keV.

At 1 keV (Fig. 2) the calculated value of the $S$-factor is 0.95 eV b, and calculation results at energies less than 50 keV are slightly lower than data of Ref. 10. In this work, using parameterization of the form

$$S(E_{c.m.}) = S_0 + E_{c.m.}S_1 + E^2_{c.m.}S_2, \qquad (19)$$

a value 2.0(2) keV mb was obtained for $S_0$; for the $S_1$ parameter the value was $1.6(4) \cdot 10^{-2}$ mb and for the $S_2$ the value $1.1(3)\ 10^{-4}$ mb keV$^{-1}$ was given. The results of this approximation are shown in Fig. 2 by the red solid line and are in a good agreement with the experimental data of Ref. 10. In Ref. 10 the results for $S$-factor of the $M$1 transition were also obtained, which lead to its value of 0.008(3) keV mb, that is in 250 times lower than the value $S_0 = 2.0(2)$ keV mb obtained in Ref. 10. In the model using by us the cross section of the $M$1 transition equals zero at all.

In Ref. 9 the equivalent values were $S_0 = 1.8(1.5)$ keV mb, $S_1 = 2.0(3.4) \cdot 10^{-2}$ mb and $S_2 = 1.1(1.4) \cdot 10^{-4}$ mb·keV$^{-1}$. The results of extrapolation these are given in Fig. 2 by the blue solid line. However, the linear extrapolation of the experimental data in Refs. 9,109 down to 1 keV leads to a value of $S$-factor about 0.6(4) eV b, i.e., three times less than in Ref. 10. In addition, the results in Ref. 10 have relatively large errors. In order to remove the current ambiguity in data for the $S$-factor of the proton capture on $^3$H, we need new measurements in the energy range from 5–10 up to 30–50 keV.

It is seen from Fig. 2 that at the lowest energies (in the region 1–3 keV) the calculated $S$-factor is practically independent of energy. This suggests that its value at zero energy will not differ from the value at 1 keV. Therefore the difference of the $S$-factor at 0 and 1 keV should not be more than 0.05 eV b and this can be considered as the error of determination of the calculated *S*-factor at zero energy, i.e., *S*(0) = 0.95(5) eV b.

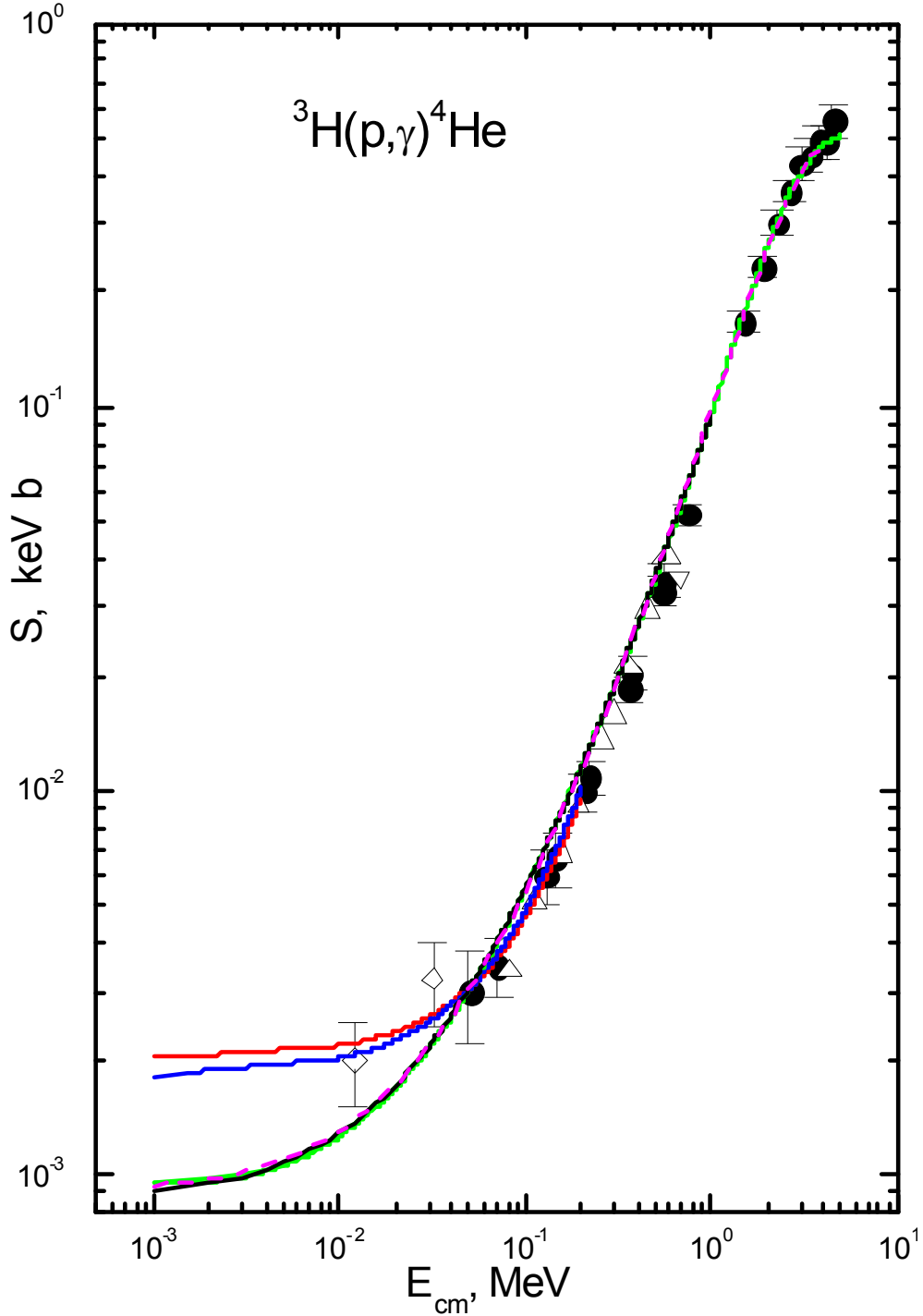

Fig. 2. Astrophysical *S*-factor of proton radiative capture on ³H in the range 1 keV–5 MeV. The green line shows the calculation with the GS ¹*S* potential given in (14); the red line shows the results of approximation from Ref. 10; the blue line shows results of the approximation from Ref. 9; the black line shows our approximation, the violet line shows our approximation with additional cubic term $E^3_{c.m.}S_3$. Points show the recalculation of the total capture cross sections Ref. 9, given in Ref. 10, upward open triangles Ref. 109, rhombus Ref. 10, downward open triangles Ref. 102.

For parameterization of the calculated *S*-factor in the energy range 1–200 keV the quadratic form (19) can be used and we obtain values for the parameters of:

$S_0 = 0.87021$ eV b, $S_1 = 4.086 \cdot 10^{-2}$ eV b keV$^{-1}$, $S_2 = 6.4244 \cdot 10^{-5}$ eV b keV$^{-2}$ at the value of $\chi^2 = 0.23$ at 1% errors of S-factor. The results of such extrapolation are shown in Fig. 2 by the black solid line. This parameterization slightly underestimates S-factor only in the range lower 2–3 keV.

It is possible to use parameterization of the form (19) with addition of the cubic term $E^3_{c.m.}S_3$. In this case at the interval up to 4 MeV for parameters we have found: $S_0 = 0.901194$ eV b, $S_1 = 3.9499 \cdot 10^{-2}$ eV b keV$^{-1}$, $S_2 = 6.94038 \cdot 10^{-5}$ eV b keV$^{-2}$, $S_3 = -1.25131 \cdot 10^{-8}$ eV b keV$^{-3}$ at the value of $\chi^2 = 0.78$ at 1% errors of calculated S-factor. The results of such extrapolation are shown in Fig. 2 by the violet dashed line. Such parameters slightly better describe the calculated S-factor at lowest energies.

For determination of the $\chi^2$ value the usual expression from Ref. 113 was used

$$\chi^2 = \frac{1}{N} \sum_{i=1}^{N} \left[ \frac{S_i^a - S_i^c}{\Delta S_i^c} \right]^2 = \frac{1}{N} \sum_{i=1}^{N} \chi_i^2, \qquad (20)$$

where $S^c$ is the initial, i.e., calculated and $S^a$ is the approximated S-factor for the energy denoted by i, $\Delta S^c$ is the error of the initial S-factor, which usually takes equal to 1%, and N is the number of points at summation in the expression given above.

In Fig. 3 the reaction rate $N_A\langle\sigma v\rangle$ of the proton capture on $^3$H is shown (solid blue line). This corresponds to the solid green line in Fig. 2 and is presented in the form Ref. 66

$$N_A\langle\sigma v\rangle = 3.7313 \cdot 10^4 \mu^{-1/2} T_9^{-3/2} \int_0^\infty \sigma(E) E \exp(-11.605 E / T_9) dE, \qquad (21)$$

where $N_A\langle\sigma v\rangle$ is the reaction rate in cm$^3$mole$^{-1}$sec$^{-1}$, E is in MeV, the cross section $\sigma(E)$ is measured in μb, μ is the reduced mass in amu and $T_9$ is the temperature in units of 10$^9$ K which matches our calculation range of 0.01 to 5.0 $T_9$. Integration of the cross sections was carried out in the range 1 keV – 5 MeV for 5000 steps with a step value of 1 keV.

In Refs. 35,38,40 the parametrization of this reaction rate is given

$$N_A\langle\sigma v\rangle = 2.20 \cdot 10^4 / T_9^{2/3} \cdot \exp(-3.869 / T_9^{1/3}) \cdot (1.0 + 0.108 \cdot T_9^{1/3} + \\ + 1.68 \cdot T_9^{2/3} + 1.26 \cdot T_9 + 0.551 \cdot T_9^{4/3} + 1.06 \cdot T_9^{5/3}) \qquad (22)$$

The calculation result of such reaction rate is shown in Fig. 3 by the dark dotted line, which appreciably differ from our results, based on the correct description of the astrophysical S-factor in the range from 50 keV to 5 MeV Ref. 91 from new work Ref. 9 – points in Fig. 2. On the analogy of (22) one can parameterize our calculation results by the analogous form (22), and with other coefficients.

$$N_A \langle \sigma v \rangle = 2.2182 \cdot 10^4 / T_9^{2/3} \cdot \exp(-4.3306/T_9^{1/3}) \cdot (1.0 + 5.7852 \cdot T_9^{1/3} - 11.374 \cdot T_9^{2/3} + 12.059 \cdot T_9) - 9.4143 \cdot T_9^{4/3} + 47.332 \cdot T_9^{5/3} \quad (23)$$

with $\chi^2 = 1.3$ at the 1% error of the parameterized reaction rate. Results of such parametrization are shown in Fig. 3 by the red dashed line.

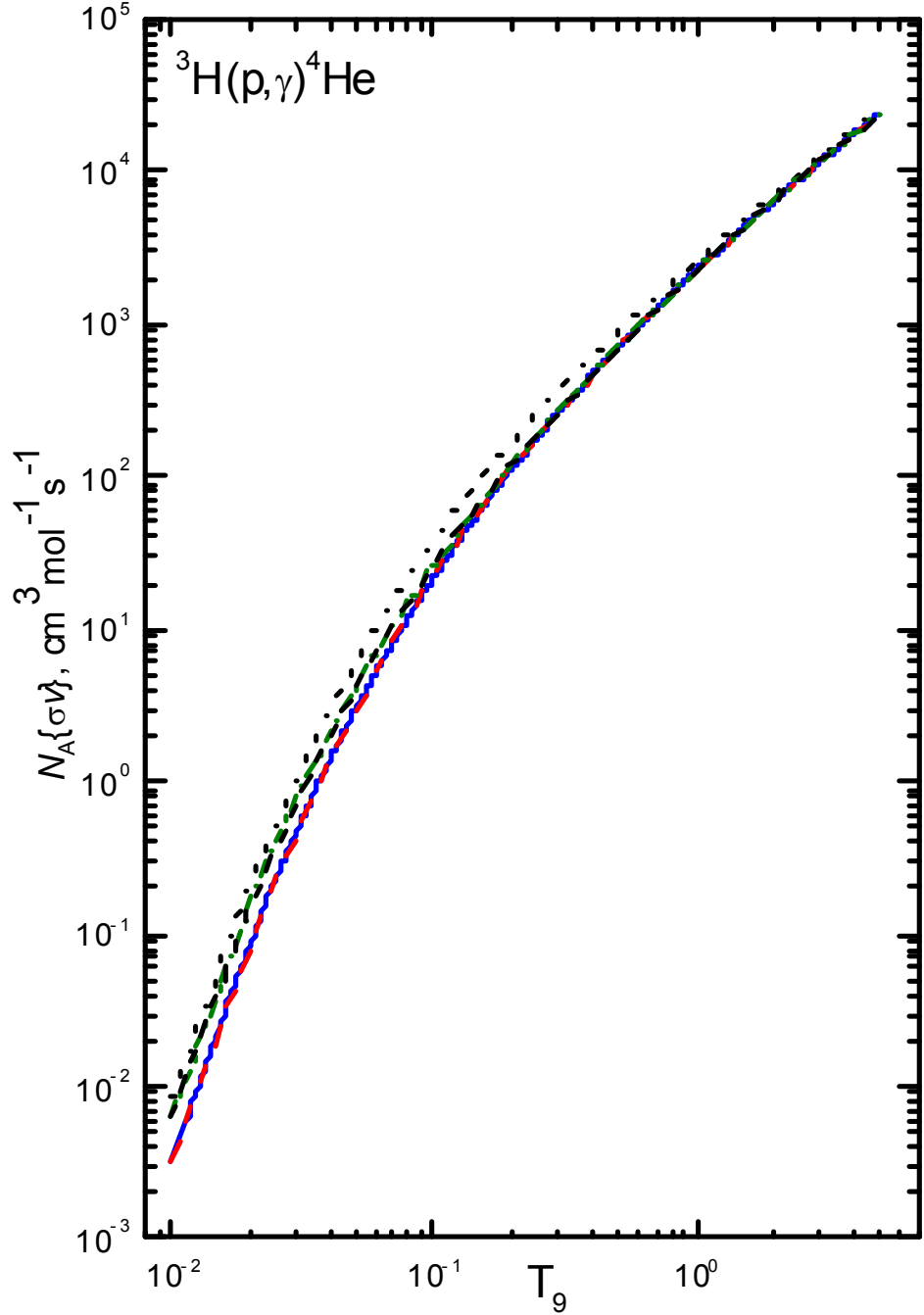

Fig. 3. Reaction rate of the proton radiative capture on $^3$H. Blue line is the calculation results for the GS potential from (14), which correspond to cross sections shown in Fig. 2 by the green solid line. Result of reaction rate (22) is shown here by the dark dotted line, reaction rate (23) is shown here by the red dashed line and the reaction rate (24) by the green dotted-dashed line.

Possibly, our results for the $S$-factors is slightly underestimated in the energy range lower 50 keV, if proceed from the results,[10] shown in Fig. 2 by open rhombus.

In order for take into account results of Ref. 10 we determine the *S*-factor from parametrization (19) with parameters of Ref. 10 and determine the cross section from the *S*-factor at the range 1 – 50 keV, which at 50 keV coincide with our calculations. Then we change our results by such cross sections in the mentioned energy range and calculate the reaction rate. The results of such calculations are shown in Fig. 3 by the green dotted-dashed line, which from 0.01 up to 0.2 – 0.3 $T_9$ is located slightly higher than our previous results shown by the blue line. However, this result, which completely describes *S*-factor in the range 1 keV – 5 MeV does not coincide with the parametrization of Ref. 38.

This result of calculations, presented in Fig. 3 by the green dotted-dashed line, can be parameterized by form (22) with parameters

$$N_A \langle \sigma v \rangle = 264.53 \cdot 10^4 / T_9^{2/3} \cdot \exp(-2.5241/T_9^{1/3}) \cdot (1.0 - 12.5629 \cdot T_9^{1/3} + \\ + 44.0654 \cdot T_9^{2/3} - 35.6503 \cdot T_9 + 59.3843 \cdot T_9^{4/3} + 49.3985 \cdot T_9^{5/3}) \quad (24)$$

with $\chi^2 = 0.1$ at 1% error of the parameterized reaction rate, which shown in Fig. 3 by the thin red solid line.

Since, there was no the most recent measurements for the astrophysical *S*-factor[9,10] at the publishing of the work Ref. 35 their parametrization slightly overestimates the reaction rate at low $T_9$, tending to our results at values 3 – 5 $T_9$. Our calculations of the reaction rate given in Fig. 3 by the green dotted-dashed line at the determination of the *S*-factor in the energy range 50 keV – 5 MeV are based on the microscopic MPCM and is not usual parametrization of the experimental data. In the range 1 – 50 keV they empirically take into account the latest data[10] at low energies and can be considered as an improvement of results.[35]

## 4. Capture reaction $^3$He($^2$H, γ)$^5$Li

The preliminary studying of the radiative capture reaction $^3$He($^2$H, γ)$^5$Li was published by us in Ref. 114. Radiative capture reactions in the isobar-analog channels $^3$H($^2$H, γ)$^5$He and $^3$He($^2$H, γ)$^5$Li are of interest for two important reasons. These reactions involving deuterium enter into the chain of synthesis of the initial formation of stable stars and are also possible candidates for overcoming a well-known problem – the gap at A = 5 in the chain of synthesis of light elements in the early stages of the evolution of the Universe.[6]

### 4.1. *Elastic $^2$H$^3$He and $^2$H$^3$H scattering and bound states*

We present the classification by orbital symmetries of the $^2$H$^3$He and $^2$H$^3$H systems, i.e. a configuration of 2+3 nucleons. The doublet channel spin (*S* = 1/2) scattering states depend on the two allowed orbital Young tableaux {41} and {32}, and these are regarded as mixed in terms of the orbital symmetries. The quartet channel spin (*S* = 3/2) allows only one symmetry {32}, so these states are pure according to the Young tableaux.

Contrary, the ground discrete states of $^5$He and $^5$Li nuclei are assumed to be the pure {41} state.[87] The different interaction potentials in the discrete and scattering states should therefore also be different relative to the symmetry of the Young tableau.

Finally, the dependence not only on *JLS* quantum numbers, but on {*f*} orbital symmetry is also taken into account for the nuclear interaction potentials of the attractive Gauss form and exponential part (2), modeling the long-range repulsion[5,20,41,43,44] with the point-like Coulomb term $V_c$ defined above.

The phase shifts of the $^2H^3He$ elastic scattering are known in the energy range 0–5 MeV.[115] Compared with the earlier research[116,117] carried out for energies of up to 40 MeV, we changed the potential parameters slightly in order to fit the low energy region better. The results are given in Table 2 for the $^2H^3He$ and $^2H^3H$ systems, as the difference is shown in the Coulomb interaction only for the same symmetry classification. For both systems, we set the repulsive part as $V_1 = 0$. The last column gives the energies of the binding states (BSs). Table 2 shows the binding energies for the quartet forbidden state with the Young tableau {5} and orbital momentum $L = 0$, as well as state with tableau {41} and $L = 1$. The allowed state corresponding to symmetry {*f*} = {32} with $L = 0$ and 2 was shown to be unbound, i.e. lying in a continuous spectrum corresponding to two exciting quanta (see Ref. 87).

Table 2. Potential parameters (2) for the $^2H^3He$ and $^2H^3H$ systems for mixed scattering states by Young tableaux and corresponding binding energies $E_{BS}$

| $^{2S+1}L$ | $V_0$, MeV | $\alpha$, fm$^{-2}$ | $E_{^2H^3He}$ ($E_{^2H^3H}$), MeV |
|---|---|---|---|
| $^2S_{1/2}$, $^2D$ | -30.0 | 0.15 | -7.0 (-7.9) |
| $^2P$ | -48.0 | 0.1 | -9.6 (-10.2) |
| $^4S_{3/2}$, $^4D$ | -34.5 | 0.1 | -13.0 (-13.9) |
| $^4P$ | -29.0 | 0.1 | -1.4 (-1.8) |

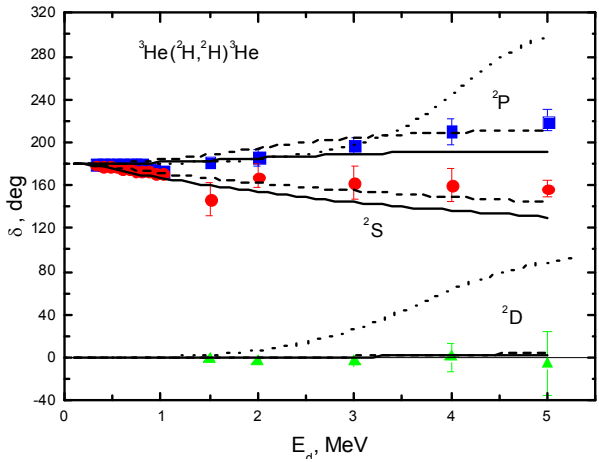
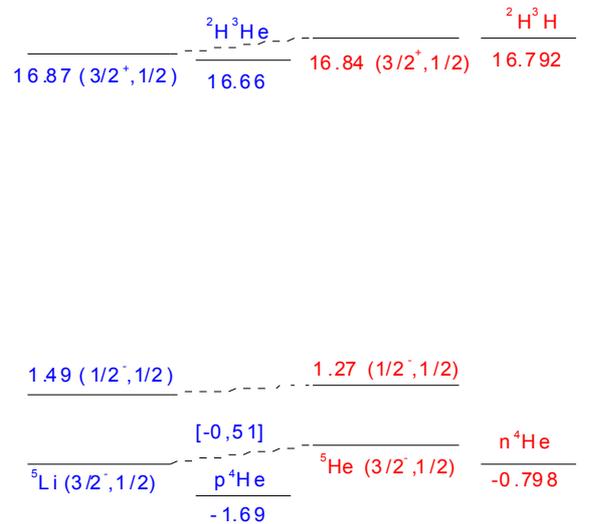

Fig. 4. Comparison of the $^2H+^3He$ doublet phase shifts mixed by orbital tableaux calculated with the potentials from Table 2 using results from Ref. 115.

Fig. 5. Experimental energy spectra for $^5Li$ and $^5He$ using data from Ref. 87.

When using the potential $V_0 = -25.0$ MeV, $\alpha = 0.15$ fm$^{-2}$ from Refs. 116,117 for the description of the doublet mixed by the Young tableaux $^2S$ phase shifts, some discrepancies from the experimental data can be seen in Fig. 4 (solid curves for $^2S$ and $^2D$ states).[115] Thus, we assume that the potential is deeper and is equal to -30.0 MeV (see Table 2). The corresponding results are shown in Fig. 4 by dashed curves. The results for the doublet $^2P$ potential -44.0 MeV, $\alpha = 0.1$ fm$^{-2}$ from Refs. 116,117 show a worse fit to the experimental data in Fig. 4 (solid curve) compared with the calculations with a deeper potential $V_0 = $ -48.0 MeV (dashed curve).

As can be seen from Table 2, in the allowed doublet $^2P$ channel, the energy of the GS and the first excited state (FES) of $^5$Li and $^5$He (experimental spectra are shown in Fig. 5) cannot be reproduced with these parameter sets. Thus, in order to fit the BSs characteristics of $^5$Li found in Refs. 116,117, the potential parameters are as given in Table 3.

Table 3. Potential parameters for $^2$H$^3$He and $^2$H$^3$H systems for pure states according to Young tableaux and binding energies $E_{BS}$ from Refs. 116,117. Here, $\alpha = 0.15$ fm$^{-2}$

| $^{2S+1}L_J$ | $V_0$, MeV | $V_1$, MeV | $\beta$, fm$^{-1}$ | $E_{^2H^3He}$ ($E_{^2H^3H}$), MeV |
|---|---|---|---|---|
| $S_{1/2}$, D | -40.0 | +8.0 | 0.2 | -8.7 (-9.6) |
| $P_{3/2}$ | -75.5 | – | – | -16.5 (-17.3) |
| $P_{1/2}$ | -60.2 | – | – | -9.0 (-9.7) |

It should be noted that both channel spin states $S = 1/2$ and $3/2$ allow the bound states in $^5$He or $^5$Li with total momentum $J = 3/2^-$ (GS) as well as $J = 1/2^-$ (FES) corresponding to the P wave. Thus, both of these states are the $^{2+4}P$ mixture of singlet and quartet channels. At one time, the pure doublet state corresponds to the {41} tableau, and the pure quartet to {32}. Thus, the $^{2+4}P$ state can also be treated as a mixture using Young tableaux. We see the direct correspondence of "channel spin" and the "Young tableau"; thus, the obtained potentials for the GS and FES with $J = 3/2^-$ and $J = 1/2^-$ are referred to here as pure using Young tableau potentials.

Based on the latest data for the energy levels[118] and refined values for the masses of the clusters involved (see Section 2) we revised the potential parameters; it can be seen from Table 4 that the binding energies in GS and FES of $^5$Li have been reproduced with a more accurate search relative precision of 10$^{-6}$ MeV.[20] It should be noted that the repulsive potential was taken as $V_1 = 0$, and the width parameter as $\alpha = 0.18$ fm$^{-2}$ in the potential (2). Experimental values for the energy levels are given in brackets next to the calculated $E_{BS}$.[118]

The dimensionless asymptotic normalizing coefficients $C_W$ are given in the last column in Table 4. They are defined above according to (3).[68] Here, $\chi_L(R)$ is the numerical GS radial WF with respect to the solution of the Schrödinger equation normalized to unity, $W_{-\eta L+1/2}(2k_0R)$ is the Whittaker function, and $k_0$ is the wave number related to the channel binding energy. The ANC error is determined by its average over the intervals 5–6 and 8–10 fm. The charge radii $r_{rms}$ for the BSs of $^5$Li in the $^2$H$^3$He channel were also calculated and are given in Table 4.

Table 4. New potential parameters for $^2$H$^3$He system for the pure by Young tableau states

| $L_J$ | $V_0$, MeV | $E_{BS}$ ($E_{exp}$), MeV | $r_{rms}$, fm | $C_W$ |
|---|---|---|---|---|
| $P_{3/2}$ | -84.03570 | -16.660002 (-16.66) | 2.25 | 6.40(1) |
| $P_{1/2}$ | -81.02697 | -15.170001 (-15.17) | 2.26 | 5.83(1) |

We draw attention here to the first resonance in the $^2$H$^3$He state with width $\Gamma_{cm} = 0.959$ MeV indicated as 19.28 MeV relative to the GS, or as 2.62 MeV relative to the channel threshold and identified as the $J^\pi = 3/2^-$ state (see Table 4.3 in Ref. 118). This may be correlated to the $P_{3/2}$ wave in the doublet or quartet spin channel at 4.37 MeV (l.s.) deuteron energy. It is shown below that $M1$ transitions from the $P_{3/2}$ scattering states close to this energy have a resonating character comparing the transitions from the non-resonating $P_{1/2}$ and $P_{5/2}$ waves, which give appreciably lesser contribution.

The phase shift analyses carried out in Ref. 115 up to 5 MeV (l.s.) did not reveal any resonating behavior of $^{2+4}P$ waves, in spite of the relatively large width of the level considered. A rising tendency is only seen in the $^2P$ phase shift in the doublet channel, while the quartet $^4P$ phase shift clearly decreases. We did not treat these resonances previously, and the potentials in Table 2 are therefore unable to reproduce these. The following potential was found to represent the $^{2+4}P_{3/2}$ resonances:

$$V_0 = -1505.3 \text{ MeV and } \alpha = 2.5 \text{ fm}^{-2}. \qquad (25)$$

The calculated $P_{3/2}$ phase shift, illustrated by dots in Fig. 4, shows the resonance. This is equal to 90.0(1)°, at 4.37 MeV (l.s.) with a width $\Gamma_{cm} = 1.06$ MeV.

There are two further wide resonances with $\Gamma_{cm} = 3.28$ and 4.31 MeV located at 19.45 and 19.71 MeV relative to the GS (or 2.79 and 3.05 MeV relative to the channel threshold). These are associated with the $J^\pi = 7/2^+$ and $J^\pi = 5/2^+$ states (see Table 4.3 in Ref. 118), and may refer to $D$ phase shifts resonating at 4.65 and 5.08 MeV (l.s.) deuteron energies. The second of these resonances in the doublet and quartet channels may indicate $E1$ transitions to the GS. To reproduce such $^{2+4}D_{5/2}$ behavior, the following potential was found:

$$V_0 = -25.5745 \text{ MeV and } \alpha = 0.075 \text{ fm}^{-2}. \qquad (26)$$

The calculated $D_{5/2}$ phase shift, illustrated by dots in Fig. 4, shows the resonance. This is equal to 90.0(1)°, at 5.08 MeV (l.s.) with width $\Gamma_{cm} = 4.43$ MeV. Again, there is no such resonance in the analysis of $D$ phase shifts carried out in Ref. 115 at these energies, but there is a tendency of a moderate rise in $^4D$ phase shift at 5.0 MeV. It is clear that as a result of the analysis of these levels the phase shift analysis[115] does not take into account the location of the considered resonances with large widths and hereafter is subject to, evidently, refinement with the extension of the deuteron energy range up to 7–8 MeV.

The complete set of transition amplitudes taken into account in our calculations is given in Table 5. Transitions from the resonating waves with a major effect on the total cross sections are marked in bold. All other transitions from the non-resonating waves (ordinary type) contribute a minor effect, according to our estimation. Scattering states corresponding to the same angular momentum $J$, but mixed by channel spin, are given in the same color.

Several comments should be made on the transitions involved in the calculations of the total cross section of the $^3$He($^2$H,$\gamma$)$^5$Li reaction. Since the GS is mixed by spins, the dipole $E1$ transition should be treated as arising from the doublet and quartet $S$ and $D$ scattering states. Within the model used, it is impossible to separate the $^2P_{3/2}$ and $^4P_{3/2}$ components explicitly in GS, and we therefore use the $P_{3/2}$ function mixed by spin states obtained as the solution of the Schrodinger equation with potentials from Table 4. For the scattering states, the doublet and quartet mixed by Young tableaux are used, calculated with potentials in Table 2.

Table 5. Transitions accounting for the calculated total cross section of the deuteron radiative capture on $^3$He

| No. | $^{2S+1}L_J$, initial | Transition | $^{2S+1}L_J$, final |
|---|---|---|---|
| 1. | $^2S_{1/2}$ | $E1$ | $^2P_{3/2}$ |
| **2.** | $^4S_{3/2}$ | **$E1$** | $^4P_{3/2}$ |
| 3. | $^2D_{3/2}$ | $E1$ | $^2P_{3/2}$ |
| **4.** | $^2D_{5/2}$ | **$E1$** | $^2P_{3/2}$ |
| 5. | $^4D_{1/2}$ | $E1$ | $^4P_{3/2}$ |
| 6. | $^4D_{3/2}$ | $E1$ | $^4P_{3/2}$ |
| **7.** | $^4D_{5/2}$ | **$E1$** | $^4P_{3/2}$ |
| 8. | $^2P_{1/2}$ | $M1$ | $^2P_{3/2}$ |
| **9.** | $^2P_{3/2}$ | **$M1$** | $^2P_{3/2}$ |
| 10. | $^4P_{1/2}$ | $M1$ | $^4P_{3/2}$ |
| **11.** | $^4P_{3/2}$ | **$M1$** | $^4P_{3/2}$ |
| 12. | $^4P_{5/2}$ | $M1$ | $^4P_{3/2}$ |

The interaction potentials are corroborated by the experimental data on the elastic scattering phase shifts and energy levels spectra; thus, the WFs obtained as the solutions of the Schrodinger equation with these potentials effectively account for the cluster system states, and in particular for the mixing by channel spin. Therefore, the total cross section of the $E1$ transition from the mixed continuous states to spin-mixed GS may be taken as a simple doubling of the partial cross section, as each is calculated with the same functions; however, spin algebraic factors are specified for each matrix element.[5,20,41,43,44] In reality, there is only one transition from the scattering state to the GS, rather than two different $E1$ processes.

The averaging procedure concerns the transitions from the $D_{5/2}$ and $D_{3/2}$ scattering states to the $P_{3/2}$ GS of $^5$Li in the $^2$H$^3$He channel. Finally, we arrive at the following $E1$ multipole cross section:

$$\sigma_0(E1) = \sigma(^2S_{1/2} \to {}^2P_{3/2}) + \sigma(^4S_{3/2} \to {}^4P_{3/2}) + \sigma(^4D_{1/2} \to {}^4P_{3/2}) + $$
$$+ [\sigma(^2D_{3/2} \to {}^2P_{3/2}) + \sigma(^4D_{3/2} \to {}^4P_{3/2})]/2 + [\sigma(^2D_{5/2} \to {}^2P_{3/2}) + \sigma(^4D_{5/2} \to {}^4P_{3/2})]/2.$$

According to the classification in Table 5, there are also spin mixing states in scattering $P$ waves leading to the magnetic dipole $M1$ transition. Thus, the corresponding cross section is written as above for the $E1$ transition to the GS:

$$\sigma_0(M1) = \sigma(^4P_{5/2} \to {}^4P_{3/2}) + [\sigma(^2P_{1/2} \to {}^2P_{3/2}) + \sigma(^4P_{1/2} \to {}^4P_{3/2})]/2 +$$
$$+ [\sigma(^2P_{3/2} \to {}^2P_{3/2}) + \sigma(^4P_{3/2} \to {}^4P_{3/2})]/2.$$

We have therefore identified all the major transitions that may contribute to the total cross sections of the deuteron capture process on $^3$He at low energies, which are treated in this paper.

## 4.2. Total cross section, astrophysical S-factor and reaction rate

Fig. 6 shows the results of the calculated $E1$ radiative capture in the $^2H^3He$ cluster channel at energies below 5 MeV. The solid red line denotes the cross section for the $E1$ transition from the $^2S$ and $^4S$ scattering waves (potentials from Table 2) to GS $^{2+4}P_{3/2}$, defined by the interaction potential parameters from Table 4. Cross sections for the $E1$ transitions from the $^2S$ wave are of few orders suppressed as this scattering wave has non-resonant behavior.

The solid violet line in Fig. 6 denotes the cross section for the $E1$ transition to the GS from the resonating $^{2+4}D_{5/2}$ waves, calculated with the potential (26), and includes all other small-value amplitudes for the non-resonating $D$ waves listed in Table 5.

The green curve in Fig. 6 shows the contribution of $M1$ transitions from the resonating $^{2+4}P_{3/2}$ waves corresponding to the potential (25) and the non-resonating set of $P$ potentials from Table 2. Note that $M1$ transitions from non-resonant scattering $P$ waves have a significant impact on the total cross sections only at energies above 600–700 keV.

The total cross section including all transitions listed in Table 5 is shown by the blue curve in Fig. 6. It can clearly be seen that starting from energies above 600–800 keV the calculated cross section lies somewhat lower than the error bars band. It should be noted that these errors have been taken as being equal to 19%, as it was given in Ref. 31 for energy 450 keV and cross section value of 21(4) μ b at the 25 keV energy scaling error.

Fig. 7 displays the calculated astrophysical $S$-factor, which is in direct correspondence with the cross sections shown in Fig. 6. We recalculated data on the cross sections from Ref. 31 into the $S$-factor and present them here as points in this figure. At minimal energies of 185–300 keV, its value is close to 0.39(5) keV b. This value may be approximated by a trivial constant energy dependence $S(E) = S_0$ with $S_0 = 0.386$ keV b and a mean value of $\chi^2 = 0.21$. The same experimental errors of 19% were assumed for the $S$-factor. The linear parametrization at energies below 20 keV is shown by the dashed green line in Fig. 7.

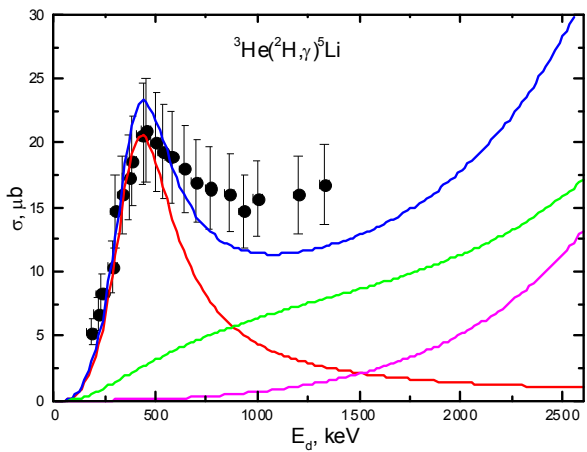 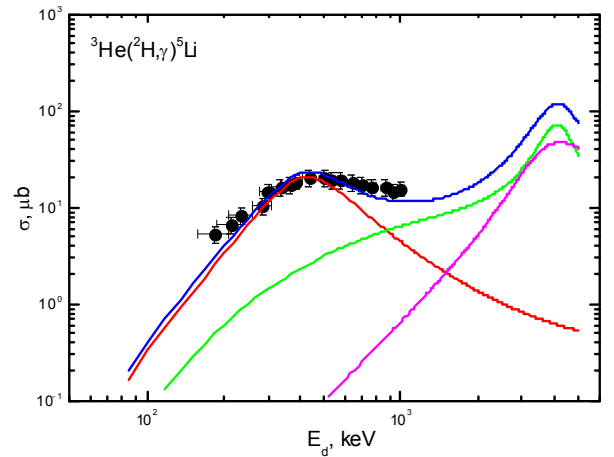

Fig. 6a. Total cross section for $^3He(^2H,\gamma)^5Li$ below 2.5 MeV. Experimental data are taken from Ref. 31 and curves correspond to the potential parameters in Tables 2 and 4

Fig. 6b. Total cross section for $^3He(^2H,\gamma)^5Li$ below 5.0 MeV. Data are the same as in Fig. 6a.

Fig. 7 displays the calculated astrophysical $S$-factor, which is in direct correspondence with the cross sections shown in Fig. 6. We recalculated data on the cross sections from Ref. 31 into the *S*-factor and present them here as points in this figure. At minimal energies of 185–300 keV, its value is close to 0.39(5) keV·b. This value may be approximated by a trivial constant energy dependence $S(E) = S_0$ with $S_0 = 0.386$ keV·b and a mean value of $\chi^2 = 0.21$. The same experimental errors of 19% were assumed for the *S*-factor. The linear parametrization at energies below 20 keV is shown by the dashed green line in Fig. 7.

To improve the description of the experimental data, we tried the following approximating function:

$$S(E) = S_0 + S_1 E + S_2 E^2. \tag{27}$$

However, this was not successful in this very low-energy region. The obtained *S*-factor was shown to be rather stable in the energy range 20–50 keV and is equal to 0.14(1) keV·b; this is substantially less than the experimental values in Ref. 31. The error given here for the calculated *S*-factor was defined as its average over the energy interval 20–50 keV.

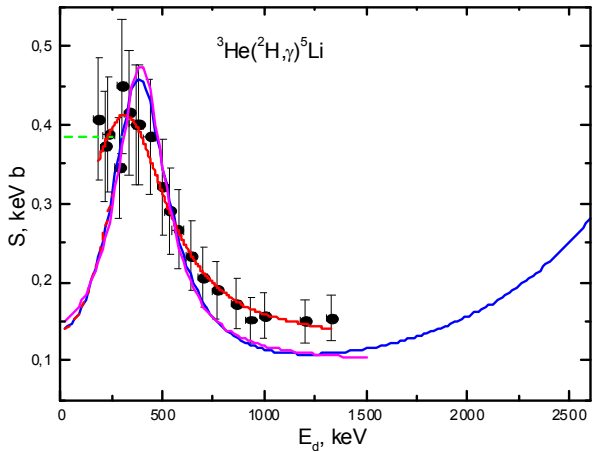 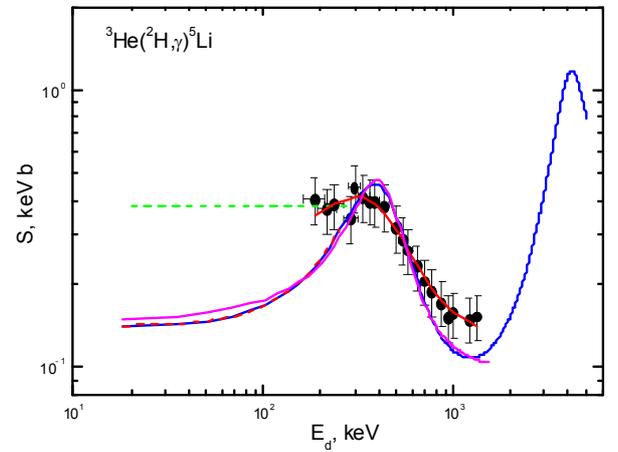

Fig. 7a. *S*-factor data from Ref. 31 fitted with potentials from Tables 2 and 4 for $^3$He($^2$H,$\gamma$)$^5$Li below 2.5 MeV

Fig. 7b. As Fig. 7a, but for energies below 5.0 MeV

In the following, we implemented the parametrization of the calculated *S*-factor according to expression (27), with $S_0 = 0.14081$ keV·b, $S_1 = -2.0705 \cdot 10^{-5}$ b and $S_2 = 2.7897 \cdot 10^{-6}$ b·keV$^{-1}$. We found the value $\chi^2$ to be 0.17 within 1% precision of the theoretical *S*-factor. The result is shown by the red dashed curve in Fig. 7 and is consistent with the experimental data in the energy region close to 250 keV.

Ordinary $\chi^2$ statistics were applied as in Refs. 41,57, and were defined as:

$$\chi^2 = \frac{1}{N} \sum_{i=1}^{N} \left[ \frac{S^a(E_i) - S^c(E_i)}{\Delta S^c(E_i)} \right]^2 = \frac{1}{N} \sum_{i=1}^{N} \chi_i^2.$$

Here, $S^c(E_i)$ are the data for the calculated astrophysical *S*-factor corresponding to the blue curve in Fig. 7. Data for the approximating value $S^a(E_i)$ are taken according to (27). The error $\Delta S^c(E_i)$ is assumed to be 1%, and *N* is the number of points taken into account.

The experimental data shown as dots in Fig. 7 may be approximated by a function of the Breit-Wigner type:

$$S(E) = a_1 + \frac{a_2}{(E - a_3)^2 + a_4^2/4}$$

with the following parameters: $a_1 = 0.12315$, $a_2 = 18861$, $a_3 = 313.22$, $a_4 = 509.95$. The results of this parametrization are shown by the solid red curve in Fig. 7. $\chi^2$ is equal to 0.11. This very approximation form was used in the energy interval up to 1.5 MeV, but with the parameter set $a_1 = 0.096846$, $a_2 = 8520.4$, $a_3 = 393.13$, $a_4 = 300.34$. The quality of the fit with $\chi^2 = 14.5$ and errors within 1% is illustrated by the solid violet curve in Fig. 7.

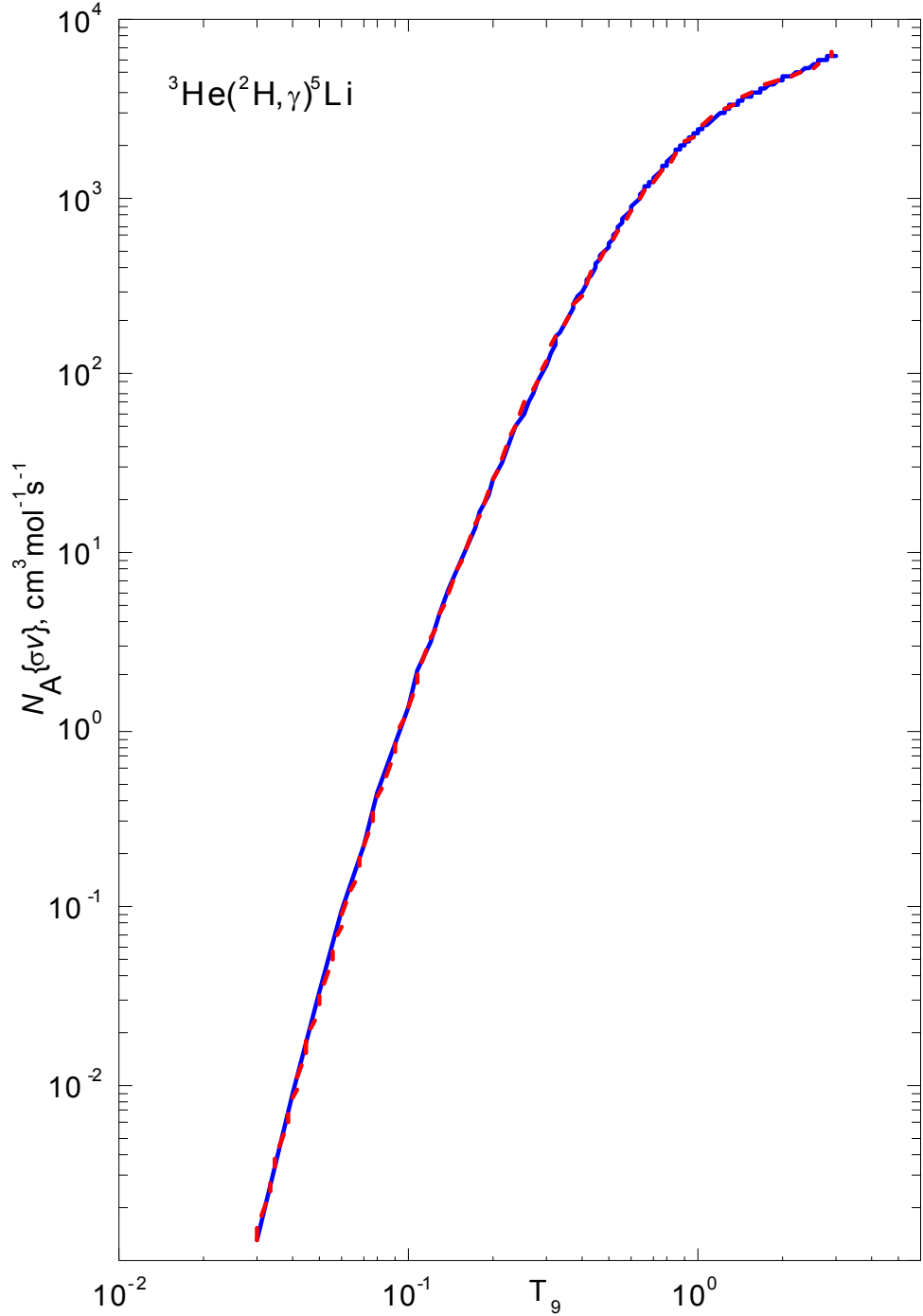

Fig. 8. Reaction rate of the deuteron radiative capture on $^3$He. Theoretical curves were obtained using the potentials from Tables 2 and 4. An explanation of the parameterization functions is given in Section 4.

Note that the value of the approximated *S*-factor at 20 keV is equal to 0.15 keV·b. Fig. 8 shows the calculated reaction rate for the deuterium radiative capture on $^3$He for temperatures from 0.03 to 3 $T_9$. The blue curve was obtained based on the corresponding theoretical cross section given in Fig. 6. The last one differs slightly from the experimental data; however, we found no measured cross sections at higher energies, at least up to 5 MeV,[60] and may be any calculated rates by other authors.

The reaction rate in cm$^3$mol$^{-1}$s$^{-1}$ units may be presented as usual (22).[66] To calculate this integral, 2000 points of the theoretical cross section were taken in the c.m. energy range from 1 to 2000 keV. The expansion of this interval to 3 MeV and number of points to 3000 changes the value of the reaction rate by less than 1%. The calculated reaction rate was approximated in the range 0.03–3.0 $T_9$ as the following:

$$N_A \langle \sigma v \rangle = 8.2838 \cdot 10^5 / T_9^{2/3} \cdot \exp(-6.6685 / T_9^{1/3}) \cdot (1.0 - 3.6550 \cdot T_9^{1/3} - 1.2850 \cdot T_9^{2/3} + 28.989 \cdot T_9 - 33.646 \cdot T_9^{4/3} + 10.779 \cdot T_9^{5/3}) \quad (28)$$

The resulting curve is shown by red in Fig. 8, where $\chi^2$ is equal to 7.2. To find the parameters in (28), we used 300 points corresponding to the blue curve in the same figure. To estimate $\chi^2$, the error was taken to be 1%.

The main goal of the present research is to determine the role of the radiative capture reaction $^3$He($^2$H,$\gamma$)$^5$Li in the balance of the processes with the deuterons occurring in the laboratory and natural plasma. This was based on the data in Ref. 35 for the parametrization of reaction rates involving the lightest and light nuclei. The contribution of the current work to this compilation is important from a practical application ansatz, and is the calculated reaction rate for the process $^3$He($^2$H,$\gamma$)$^5$Li and its analytical parametrization in (28). Fig. 1 displays the comparative rates of the deuteron- and proton-induced reactions calculated according to the parametrizations in Refs. 16,35 and the present results. It is obvious that the treated reaction is nearly four orders of magnitude smaller than the $^3$H(d,n)$^4$He process that contributes the primary effect of the primordial nucleosynthesis of the lightest elements in the Universe together with the reactions $^3$He(d,p)$^4$He, $^2$H(d,n)$^3$He and $^2$H(d,p)$^3$H.

## 5. Capture reactions $^4$He($^3$He, $\gamma$)$^7$Be, $^4$He($^3$H, $\gamma$)$^7$Li and $^4$He($^2$H, $\gamma$)$^6$Li

### 5.1. Potentials and scattering phase shifts

It was shown in Ref. 119 that orbital states in the $^3$He$^4$He, $^3$H$^4$He and $^2$H$^4$He systems for $^7$Be, $^7$Li and $^6$Li nuclei, unlike lighter cluster systems as p$^2$H or p$^3$H,[42,88,120–122] are pure according to Young tableaux. Therefore, nuclear potentials of the form (2) with the obtained parameters based on quantum numbers *JLS* and Young tableaux {*f*} and obtained on the basis of elastic scattering phase shifts and spherical or point-like Coulomb term[113] can be directly used for examination of the characteristics of the bound states (BSs) of these nuclei in the potential cluster model with FSs, which is titled as MPCM.[57]

The parameters of Gaussian interaction potentials for cluster states in $^7$Li, $^7$Be and $^6$Li nuclei that are pure according to Young tableaux[119] obtained earlier in our works,[73,123–125] meanwhile the interactions in the $^3$H$^4$He and $^3$He$^4$He systems differ from each other by the Coulomb term only. Furthermore we will give few variants of parameters, adjusted later and initial variants are given in Table 6. Table 6 also presents the energies of bound forbidden states for the $^2$H$^4$He channel of $^6$Li and the $^3$H$^4$He system, which slightly differ from the corresponding values for $^3$He$^4$He interactions. In the $S$ wave for the $^3$H$^4$He and $^3$He$^4$He systems these bound states correspond to forbidden Young tableaux {7} and {52}. In the $P$ wave the {61} tableau is forbidden at the allowed bound state (AS) with the Young tableau {43}. In the $D$ wave there is the FS with tableau {52}.[119–121,126,127] For the $^2$H$^4$He system, the $S$ wave contains the forbidden bound state with tableau {6} and allowed bound state with {42}, and in the $P$ wave the state with tableau {51} is forbidden.[119–121,126,127]

Table 6. Potential parameters of the elastic $^3$H$^4$He, $^3$He$^4$He and $^2$He$^4$He scattering and energies of corresponding forbidden bound states.[73,123–125,127] The potential width parameter for the $^3$H$^4$He and $^3$He$^4$He systems is α = 0.15747 fm$^{-2}$, and the Coulomb radius $R_{Coul}$ = 3.095 fm. $R_{Coul}$ = 0 for potentials of the $^2$He$^4$He system

| | $^7$Li and $^7$Be | | | $^6$Li | | |
|---|---|---|---|---|---|---|
| $^{2S+1}L_J$ | $V_0$, (MeV) | $E_{FS}$ ($^7$Li) (MeV) | $^{2S+1}L_J$ | $V_0$ (MeV) | α (fm$^{-2}$) | $E_{FS}$ (MeV) |
| $^2S_{1/2}$ | -67.5 | -36.0, -7.4 | $^3S_1$ | -76.12 | 0.2 | -33.2 |
| $^2P_{1/2}$ | -81.92 | -27.5 | $^3P_0$ | -68.0 | 0.22 | -7.0 |
| $^2P_{3/2}$ | -83.83 | -28.4 | $^3P_1$ | -79.0 | 0.22 | -11.7 |
| $^2D_{3/2}$ | -66.0 | -2.9 | $^3P_2$ | -85.0 | 0.22 | -14.5 |
| $^2D_{5/2}$ | -69.0 | -4.1 | $^3D_1$ | -63.0 (-45.0) | 0.19 (0.15) | – |
| $F_{5/2}$ | -75.9 | – | $^3D_2$ | -69.0 (-52.0) | 0.19 (0.15) | – |
| $F_{7/2}$ | -84.8 | – | $^3D_3$ | -80.88 | 0.19 | – |

The quality of description of scattering phase shifts is demonstrated in Figs. 9, 10 and 11a,b,c; these figures also show experimental data from Refs. 128,129 for the $^3$He$^4$He, Refs. 129,130 for the $^3$H$^4$He, and Refs. 131–134 for the $^2$H$^4$He elastic scattering.

For the $^3$He$^4$He and $^3$H$^4$He systems, only the $S$ scattering phase shifts are presented, since as it will be shown below, exactly transitions from the $S$ waves to the ground and the first excited bound states of $^7$Be and $^7$Li make a dominant contribution to the radiative capture $S$-factor. It can be seen from Figs. 9, 10, and 11a that the calculated $S$ phase shifts for the elastic $^3$H$^4$He, $^3$He$^4$He and $^2$H$^4$He scattering quite well describe the known results of the phase shift analysis at low energies up to 5–10 MeV. Fig. 11b shows that the data on the $P$ phase shifts of the $^2$H$^4$He scattering from different papers strongly differs; therefore, it is possible to construct the $P$ potential only approximately; however, on the whole they describe phase shifts at low energies, representing a certain compromise between the results of different phase shift analyses. For energy below 1 MeV, i.e., in the region of determination of the $S$-factors, the results of calculation of all $P$ phase shifts slightly differ and are close to zero. In Fig. 11c dashed lines show the results of calculation of the $^3D_1$ and $^3D_2$ phase shifts of the $^2$H$^4$He scattering obtained using the changed potentials, the parameters of which are given in Table 6. These potentials somewhat better describe the behavior of the available experimental data, especially at the energies higher than 5 MeV.

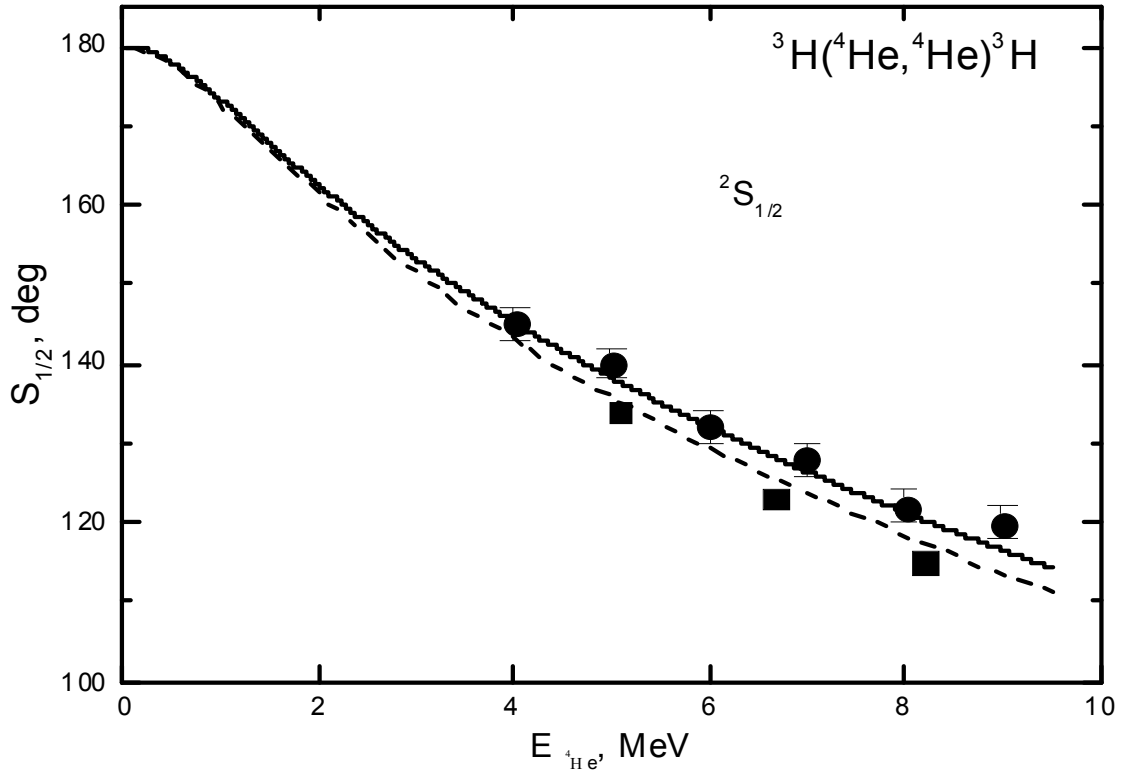

Fig. 9. $^2S_{1/2}$ phase shift of the elastic $^3$H$^4$He scattering at low energies. Experimental data from Ref. 129 is shown by filled circles and data from Ref. 130 by filled squares. The dashed line shows results for the second variant of the $S$-potential from Table 6.

It should be note that all $S$ scattering phase shifts at zero energy are shown in Figs. 9, 10, and 11a beginning from a value of 180°, although in the presence of two bound (allowed or forbidden) states in all systems, according to the generalized Levinson theorem,[119] they should start from 360°. Fig. 11b shows the $P$ phase shifts of $^2$H$^4$He scattering at zero energy beginning from 0°, although, in the presence of a bound forbidden state with scheme {51}, they should be started from 180°.

Then intercluster interactions thus matched with scattering phase shifts were used for calculation of different characteristics of the ground states of $^7$Li, $^7$Be, and $^6$Li and electromagnetic processes in these nuclei, and clusters were matched with corresponding properties of free nuclei.[56,126] The parameters of the potentials of the ground states in the $P$ wave for the $^3$He$^4$He and $^3$H$^4$He systems, and the $S$ wave for the $^2$H$^4$He were determined first of all based on correct description of binding energy.[56] In the latter case, it is possible to reproduce not only the binding energy, but also to correctly reproduce the behavior of the $S$ phase shift of elastic scattering at low energies (Fig. 11a). It should be noted that all results were obtained without taking into account tensor forces.[135]

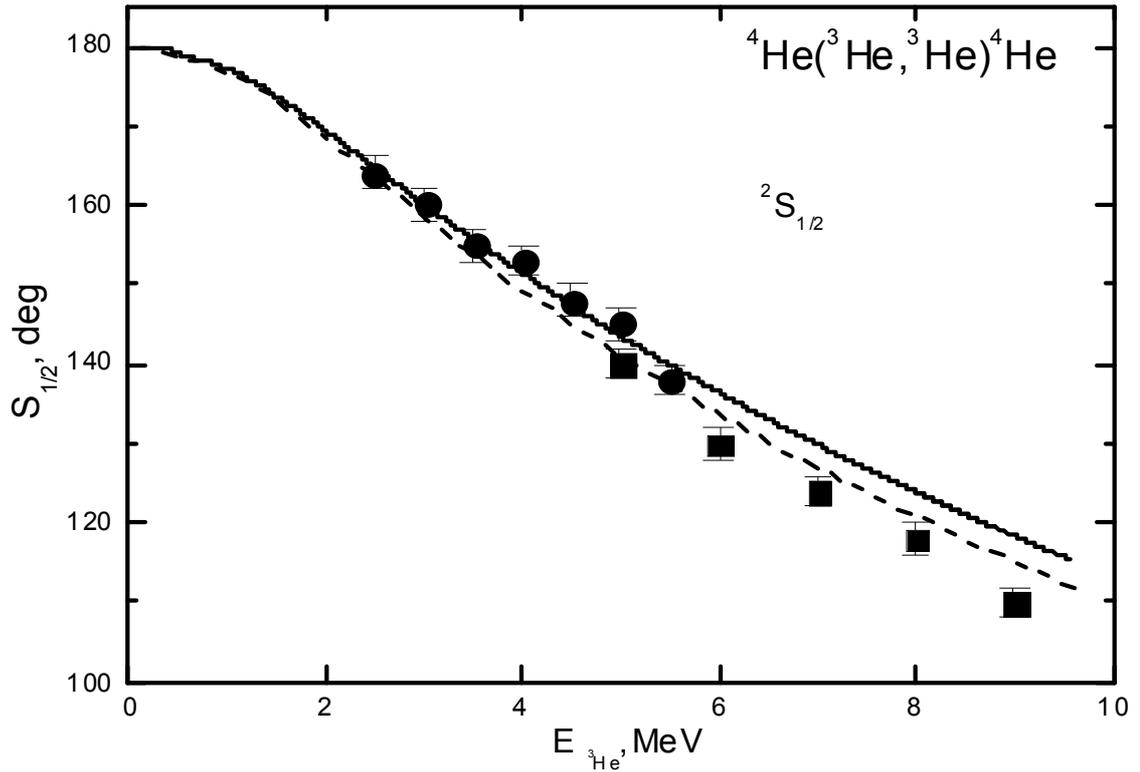

Fig. 10. $^2S_{1/2}$ phase shift of the elastic $^3$He$^4$He scattering at low energies. Experimental data from Ref. 128 is shown by filled circles and data from Ref. 129 by filled squares. The dashed line shows results for the second variant of the *S*-potential from Table 6.

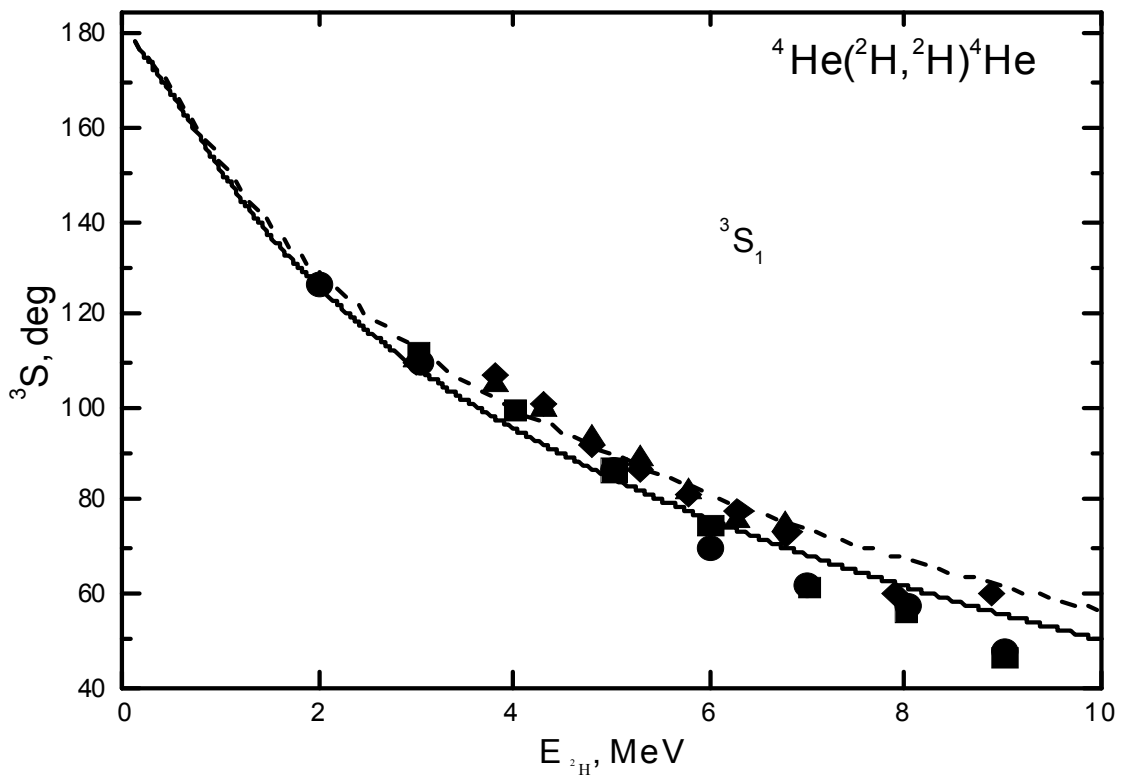

Fig. 11a. $^3S_1$ phase shift of the elastic $^2$H$^4$He scattering at low energies. Experimental data from Ref. 131 is shown by filled circles, Ref. 132 filled squares, Ref. 133 filled triangles, and Ref. 134 filled rhombs.

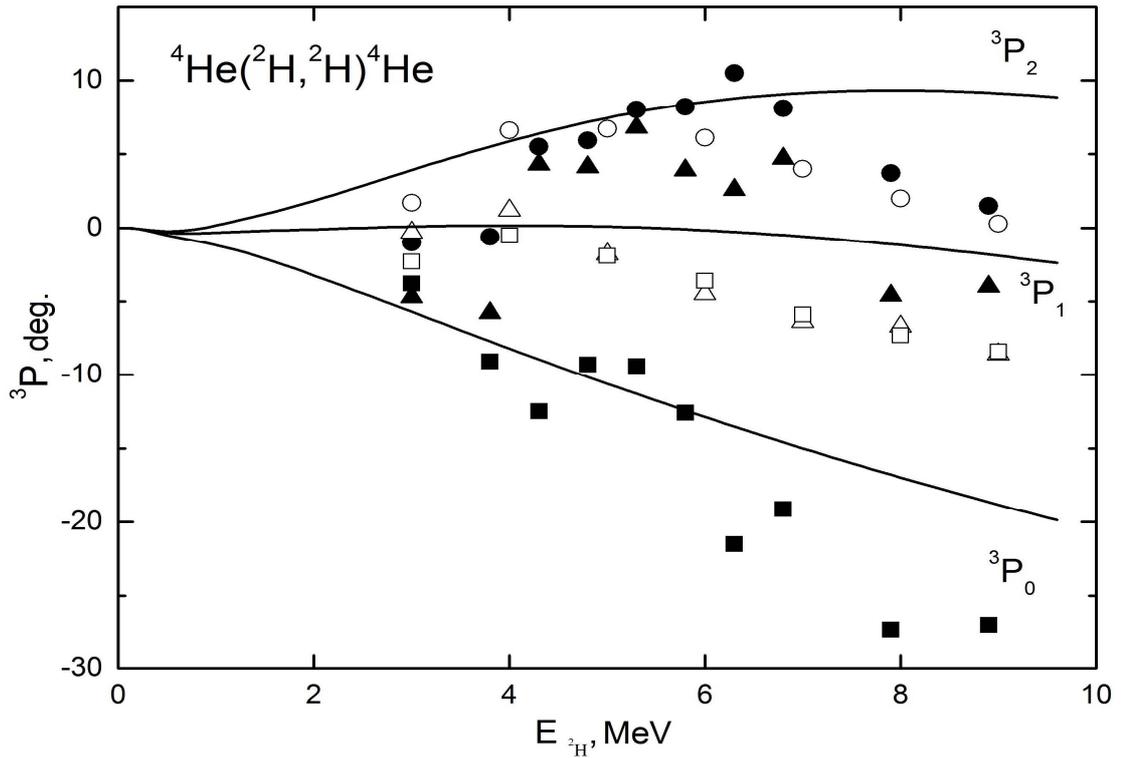

Fig. 11b. $^3P$ phase shift of the elastic $^2H^4He$ scattering at low energies. Experimental data from Ref. 132 is shown by open circles $P_2$, open triangles – $P_1$, and open squares – $P_0$; data from Ref. 134: points – $P_2$, triangles – $P_1$, and squares – $P_0$.

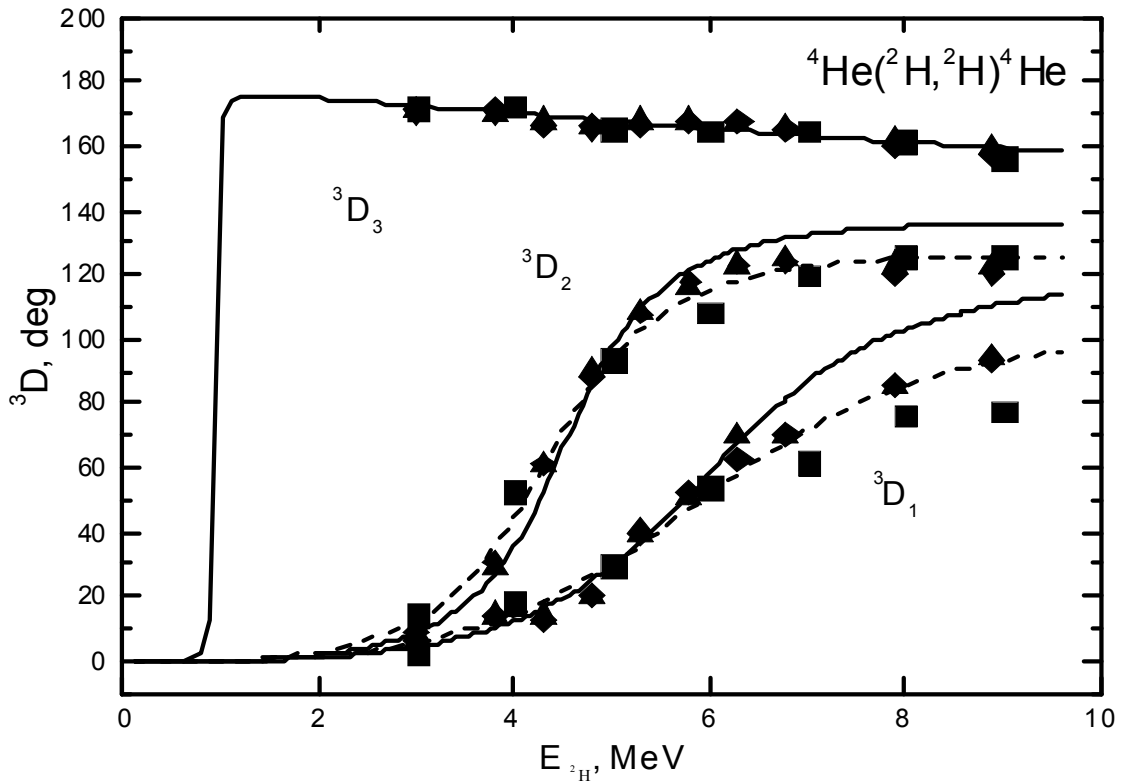

Fig. 11c. $^3D$ phase shifts of the elastic $^2H^4He$ scattering at low energies. Experimental data from Ref. 132 is shown by filled squares, Ref. 133 filled triangles, and Ref. 134 filled rhombs.

## 5.2. New variants of potentials

In the framework of this approach, the good agreement of calculations with different experimental data was obtained for both electromagnetic processes and the main characteristics of bound states of many light nuclei in cluster channels.[73,123–126] However, for example, the binding energy of $^7$Li in the $^3$H$^4$He channel with $J = 3/2^-$, as well as other systems, was really determined with an accuracy of several kiloelectronvolts; therefore, the accuracy of calculation of the S-factor of the radiative capture even at 10 keV turned out to be relatively low.

Therefore, the main calculated characteristics of bound states for $^7$Li, $^7$Be, and $^6$Li nuclei in the $^3$H$^4$He, $^3$He$^4$He, and $^2$H$^4$He channels have been refined in Refs. 136,137. For this purpose the parameters of the potentials for bound states have been refined, the calculated energy levels completely describe the experimental values.[118] In other words, the potential parameters were chosen in such a way that experimental energy levels were described with maximal possible accuracy. The energies of the bound levels of the considered nuclei for the given potentials were calculated using the finite-difference method (FDM)[138] with an accuracy no worse than $10^{-6}$ MeV. New parameters of the $^3$H$^4$He, $^3$He$^4$He and $^2$H$^4$He BS potentials are given in Table 7.

Table 7. Refined potential parameters of the $^3$H$^4$He, $^3$He$^4$He and $^2$H$^4$He interactions; energy levels; and charge radii of $^7$Li, $^7$Be and $^6$Li nuclei calculated with these potentials. The parameter $\alpha$ for $^3$H$^4$He and $^3$He$^4$He systems is 0.15747 fm$^{-2}$, and $R_{\text{Coul}} = 3.095$ fm. For $^2$H$^4$He scattering, it is assumed that $R_{\text{Coul}} = 0$ fm

| | $^7$Li | | | $^7$Be | | |
|---|---|---|---|---|---|---|
| $^{2S+1}L_J$ | $V_0$ (MeV) | $E$ (MeV) | $<r^2>^{1/2}$ (fm) | $V_0$ (MeV) | $E$ (MeV) | $<r^2>^{1/2}$ (fm) |
| $^2P_{3/2}$ | -83.616808 | -2.467000 | 2.46 | -83.589554 (-75.503718 $\alpha = 0.14$) | -1.586600 | 2.64 (2.70) |
| $^2P_{1/2}$ | -81.708413 | -1.990390 | 2.50 | -81.815179 | -1.160820 | 2.69 |
| | $^6$Li | | | | | |
| $^{2S+1}L_J$ | $V_0$ (MeV) | $\alpha$ (fm$^{-2}$) | $E$ (MeV) | $<r^2>^{1/2}$ (fm) | | |
| $^3S_1$ | -75.8469155 (-92.07748) | 0.2 (0.25) | -1.474300 | 2.65 (2.58) | | |

The slight change of parameters of potentials with respect to results of our previous works[73,74,123–126] and Table 6 practically does not influence the behavior of the scattering phase shifts. However, this change makes it possible to maximally accurately reproduce energy levels in cluster channels. The potential widths in Tables 6, 7 were chosen from the condition of description of charge radii and asymptotic constants.[136,137] Table 7 presents the results of calculation of the charge radii of considered nuclei in the cluster channels. For finding the charge radius of the nucleus, cluster radii given in Ref. 74 were used; the values of these radii, together with the energies of the bound states in cluster channels and cluster masses, are given in Tables 8.

Table 8. Experimental data on charge radii[139] and binding energies[118]

| Nucleus and its binding energy in the given channel (MeV) | Radius, (fm) |
|---|---|
| $^2$H | 2.1415(86) |
| $^3$H | 1.7591(363) |
| $^3$He | 1.9664(23); 1.9659(30) |
| $^4$He | 1.6753(28) |
| $^6$Li<br>$E(^4$He$^2$H$) = -1.4743$ | 2.546(3) |
| $^7$Li<br>$E(^4$He$^3$H$) = -2.467\ (3/2^-); -1.99039\ (1/2^-)$ | 2.410(8) |
| $^7$Be<br>$E(^4$He$^3$He$) = -1.5866\ (3/2^-); -1.16082\ (1/2^-)$ | - |

For examination of the stability of the "tail" of the wave function of ground and first excited bound states at large distances, the dimensionless asymptotic constant (AC) $C_W$ has been used. We use expression (3) where $\chi_L(R)$ is the numerical wave function of the bound state obtained from the solution of the radial Schrödinger equation and normalized to unity; $W_{-\eta L+1/2}$ is the Whittaker function of the bound state determining the asymptotic behavior of the wave function which is the solution to the same equation without the nuclear potential; i.e., at large distances $R$; $k_0$ is the wave number caused by the channel binding energy of systems; $\eta$ is the Coulomb parameter; and $L$ is the orbital angular moment of the bound state.

As a result, for AC of ground states of $^7$Be, $^7$Li, and $^6$Li nuclei, in the considered channels, the following values were obtained: 5.03(1), 3.92(1), and 3.22(1), respectively. For the first excited states (FES) of $^7$Be and $^7$Li nuclei, in this model the following values were found: 4.64(1) and 3.43(1). The error given in brackets is determined by averaging the constant for nuclei with $A = 7$ obtained in calculation on an interval of 6–16 fm and for $^6$Li at an interval of 5–19 fm.

Let us give small review of these ACs, obtained in other works. The value 2.93(15) was obtained from the $^2$H$^4$He scattering elastic phase shifts in Ref. 140. The values from 2.09 to 3.54 are obtained after recalculation to the dimensionless quantity at $k_0 = 0.308$ fm$^{-1}$ in Ref. 141 for different types of NN and N$\alpha$ interactions. At the same work there are references to experimental data that after the same recalculation have values from 2.92(25) to 2.96(14). The value of 3.04 for dimensionless AC was obtained in the earlier work.[142] We can see from here that the AC of the GS of $^6$Li usually has the value about three, and potential from Table 7 leads to slightly large value. Because for the GS potential of $^6$Li both radius and AC are inflated, we slightly change here potential parameters, having done it narrower – these values are given in Table 7 in brackets. It leads to the AC equals 2.93(1) at the range of 5–23 fm, and the radius decreased down to 2.58 fm – both these values better agree with data from Table 8 and Refs. 140,141. The phase shifts of this potential are shown in Fig. 11a by the dashed line, which is in a better agreement with the results of phase shifts analysis.[134]

The values of the asymptotic constant for the ground state of the $^3$H$^4$He system in $^7$Li equals 3.88(16) and for the FES equals 3.22(15) were obtained in Ref. 143 using recalculation to the dimensionless quantity with $k_0 = 0.453$ fm$^{-1}$. The value of 3.74(26) was given in Ref. 144 for the GS after recalculation to the dimensionless quantity. The value of 5.66(16) that is slightly more than our calculated value, is proposed for the ground state of $^7$Be in the $^3$He$^4$He channel in Ref. 145 on the basis of different experimental data after using recalculation to the dimensionless quantity with $k_0 = 0.363$ fm$^{-1}$. The value 4.66(15) is given for the first excited state, that is in a good agreement with the value obtained here. We also slightly change parameters for the GS potential of $^7$Be – these values are given in Table 7 in brackets. The new potential leads to the AC equals of 5.57(1) at the range of 6–18 fm, which is in a better agreement with results of Refs. 143,144.

### 5.3. Astrophysical S-factors

Earlier total cross sections of photoprocesses and astrophysical *S*-factors for the $^2$H$^4$He, $^3$H$^4$He and $^3$He$^4$He systems were calculated in a cluster model[146] similar to that applied here and also in the frame of the RGM.[147] For interaction potentials with forbidden states, the total photodisintegration cross sections in the $^2$H$^4$He cluster channel of $^6$Li were calculated based on three-body wave functions of the ground state.[148] The total cross sections of photoprocesses for $^6$Li and $^7$Li nuclei in two-cluster models with forbidden states were calculated in our works Refs. 73,74,123,126.

In relation with publication of new experimental data, we consider the astrophysical *S*-factors of radiative capture of $^4$He($^3$He, γ)$^7$Be, $^4$He($^3$H, γ)$^7$Li and $^4$He($^2$H, γ)$^6$Li reactions at energies as low as possible in the framework of the potential cluster model[73,74,123–126] with FSs and specified here potentials of the ground states of $^7$Li, $^7$Be, and $^6$Li nuclei (Table 7). Only the *E*1 transitions were taken into account in calculations for $^3$H$^4$He and $^3$He$^4$He systems, since the contributions of *E*2 and *M*1 transitions are lower by two to three orders of magnitude.[74] In these systems only the *E*1 transitions between the ground $P_{3/2}$ state of $^7$Li, $^7$Be, and $S_{1/2}$, $D_{3/2}$, $D_{5/2}$ scattering states and between the first excited bound $P_{1/2}$ state and $S_{1/2}$, $D_{3/2}$ scattering states are possible.

### 5.3.1. S-factor of the $^4$He($^3$H, γ)$^7$Li capture

The results of calculation of the astrophysical *S*-factor of radiative capture of $^4$He($^3$H, γ)$^7$Li reaction at energies down to 10 keV are shown in Fig. 12a by the solid line. Experimental data are taken from;[149–152] in the first of them the results for energies down to 55 keV are presented. Our calculated value of the *S*-factor at 10 keV is equal to 106.1 eV b (earlier in Refs. 41,136,137 we have obtained 111.0 eV b) at the value extracted from experiment for zero energy was equal to 106.7(4) eV b.[149] There are other data, which result in the following values at zero energy: 140(20)[152] and 100(20) eV b.[66] Among theoretical calculations, results of cluster model[146] can be noted; in this model the value of *S*-factor equals 100 eV b was obtained at zero energy, the value in the resonating group method is equal to 98(6) eV b,[147] and in recent paper[143] $S(0) = 97.4(10)$ eV b.

The difference of our previous value 111 eV b from the present one equals less than 5% and dependent from the using in FES and GS the same GS potential. Energies of these levels slightly differ from each other and, as we see this now, the results for the *S*-factor as little as by 5%. At the same time, the experimental errors of the *S*- factor at low energy, even in Ref. 149 were they minimal, are equal about to 10%. The difference of the results for the *S*-factor at zero energy range to 40%. In the present calculation we used correct FES potential from Table 7 and thereby we specified value of the *S*-factor.

If only transitions from the *S* scattering wave to the ground and first excited bound *P* states are taken into account, at 10 keV we obtain the value of 110.8 eV b, which practically does not differ from that presented above. The results of *S*-factor calculation in this case are shown in Fig. 12a by the red dashed line. It differs from our results, which take into account transitions from *D* waves, only at energies above 0.2–0.3 MeV.

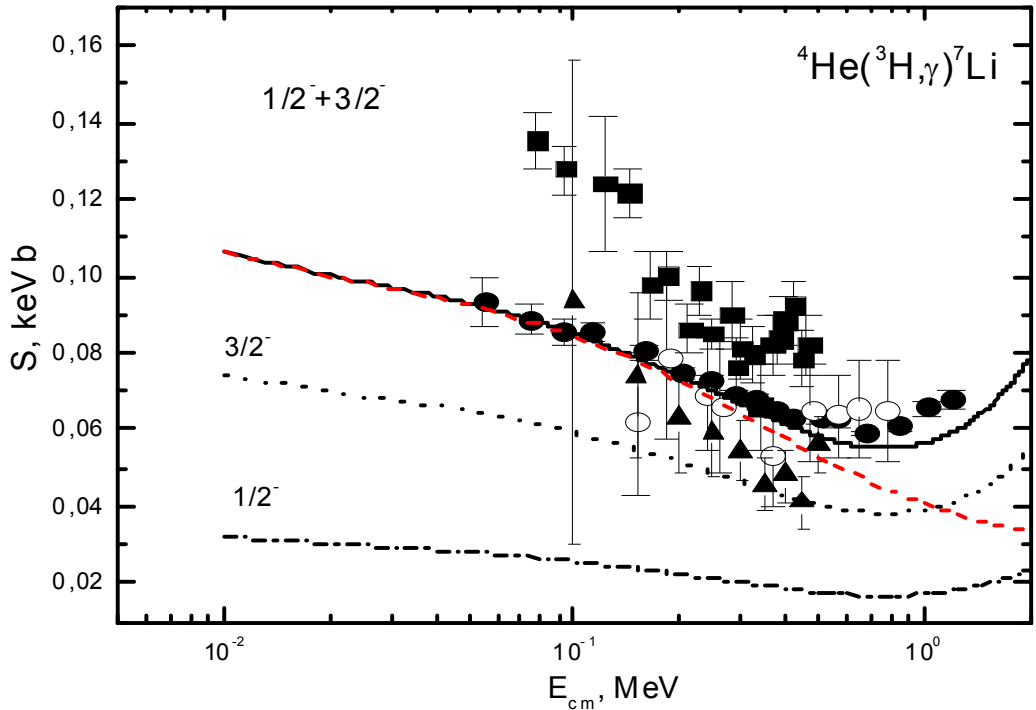

Fig. 12a. Astrophysical *S*-factor of the radiative $^4$He($^3$H, $\gamma$)$^7$Li capture. Filled circles show experimental data Ref. 149, open circles Ref. 150, and filled squares Ref. 152, triangles – from data of Ref. 151. Lines show results of calculation with parameters of GS and FES potentials from Table 7.

Thus, the results of our calculation of the *S*-factor of $^4$He($^3$H, $\gamma$)$^7$Li radiative capture quite well describe experimental data[149] at energies below 0.5 MeV. They lead to the *S*-factor value at 10 keV, which is in reasonable agreement with extrapolation of this experiment to zero energy. It should be noted that our first-ever calculations for this *S*-factor resulted in 87 eV b.[73,74,123] This difference may be related with the properties of the bound state potential used earlier in calculations (see Table 6) that yielded the binding energy with accuracy to several keV.[74]

Furthermore, the rate of the considered reaction is shown in Fig. 12b at the range from 0.05 to 5 $T_9$ is expressed by the usual way (22).[66] The capture reaction rate to the FES was shown in Fig. 12b by the dotted line, to the GS by the dashed line, and the total reaction rate by the solid line, which corresponds to the solid line in Fig. 12a. The total reaction rate shown in Fig. 12b can be approximated by the form[153]

$$N_A \langle \sigma v \rangle = 2304.319/T_9^{2/3} \cdot \exp(-6.165/T_9^{1/3}) \cdot (1.0 - 25.706 \cdot T_9^{1/3} +$$
$$74.057 \cdot T_9^{2/3} + 28.460 \cdot T_9 - 61.303 \cdot T_9^{4/3} + 19.591 \cdot T_9^{5/3}) + 29.322/T_9^{3/2} \cdot \exp(-1.641/T_9)$$

which leads to $\chi^2 = 0.4$ at 1% errors for theoretical curve. Here and then the parametrization of the calculated curve consists of 500 points was carried out.

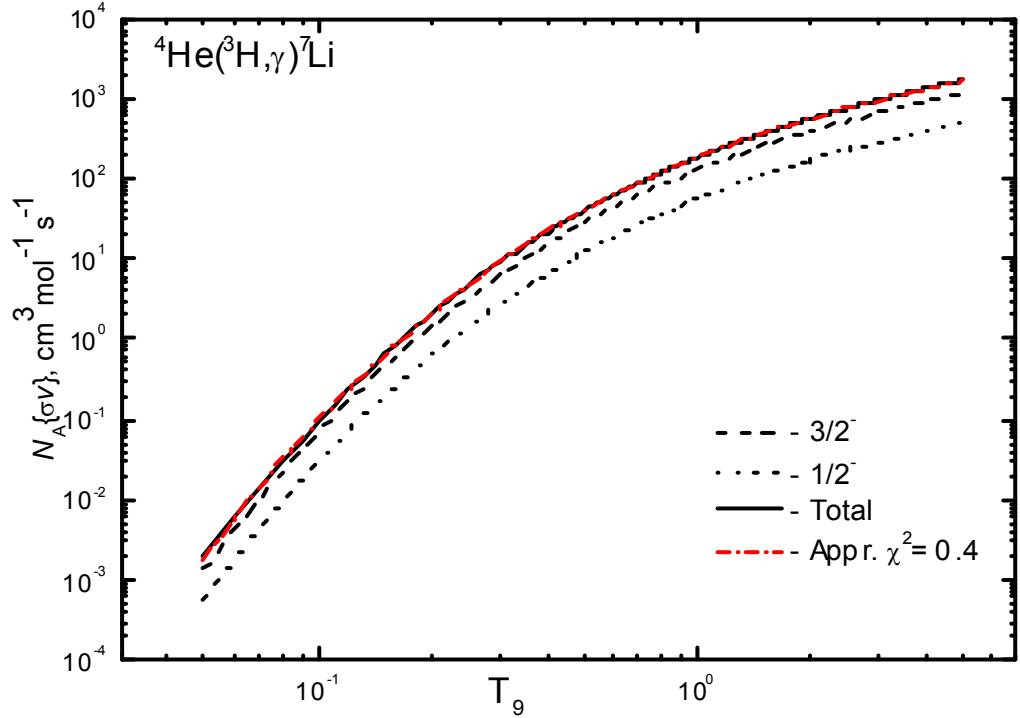

Fig. 12b. Reaction rate of the $^4$He($^3$H, $\gamma$)$^7$Li capture at the energy range from 0.05 to 5 $T_9$.

## 5.3.2. S-factor of the $^4$He($^3$He, $\gamma$)$^7$Be capture

The results of calculation of the astrophysical $S$-factor of the radiative $^4$He($^3$He, $\gamma$)$^7$Be capture at energies down to 20 keV are shown in Fig. 13a by a solid line. Experimental data has been taken from Refs. 154–160. The calculated curve at energy below 200 keV best agrees with results[155,156] obtained recently, and partially with data[158] at energies below 0.5 MeV. For energy of 20 keV, our calculation yields an $S$-factor of 0.560 keV b, and for 23 keV it was equal to 0.552. Earlier in our works[41,Ошибка! Закладка не определена.,136,137] for this value was obtained 0.593 keV b, i.e., 6% more, because for the GS and the FES one and the same GS potential was used. This is not essentially, because $S$-factor errors in the measured earlier energy range from 90 keV and higher are equal to 10–20%, and data at lower energies are absent. Now, when the new data at low energy[160] has appeared, we specified the $S$-factor value using for this the FES potential from Table 7.

For comparison we present some results of extrapolation of experimental data to zero energy: 0.54(9) keV b,[66] 0.550(12) keV b,[161] 0.595(18) keV b,[154] 0.560(17) keV b,[155] 0.550(17) keV b,[156] and 0.567(18) keV b.[162] Recently, it was obtained, in Ref. 145 based on analysis of different experimental data, that $S(0) = 0.610(37)$ keV b and $S(23 \text{ keV}) = 0.599(36)$ keV b, which well agrees with the value found here. The latest measurements of the $S$-factor at energy $23\binom{+6}{-5}$ keV Ref. 160 lead to the value 0.548(54) keV b, which absolutely agrees with our results obtained in Ref. 136 and specified here. As one can see in Fig. 13a the results of our calculations at this energy lay in the range of experimental errors of Ref. 160. Let us notice that our calculations of this $S$-factor were done in work Ref. 136 in 2010 and became a part of review Ref. 137, but new data[160] were published in 2015. Intrinsically, theoretical results of Ref. 136 predict behavior of the $S$-factor at lowest energies.

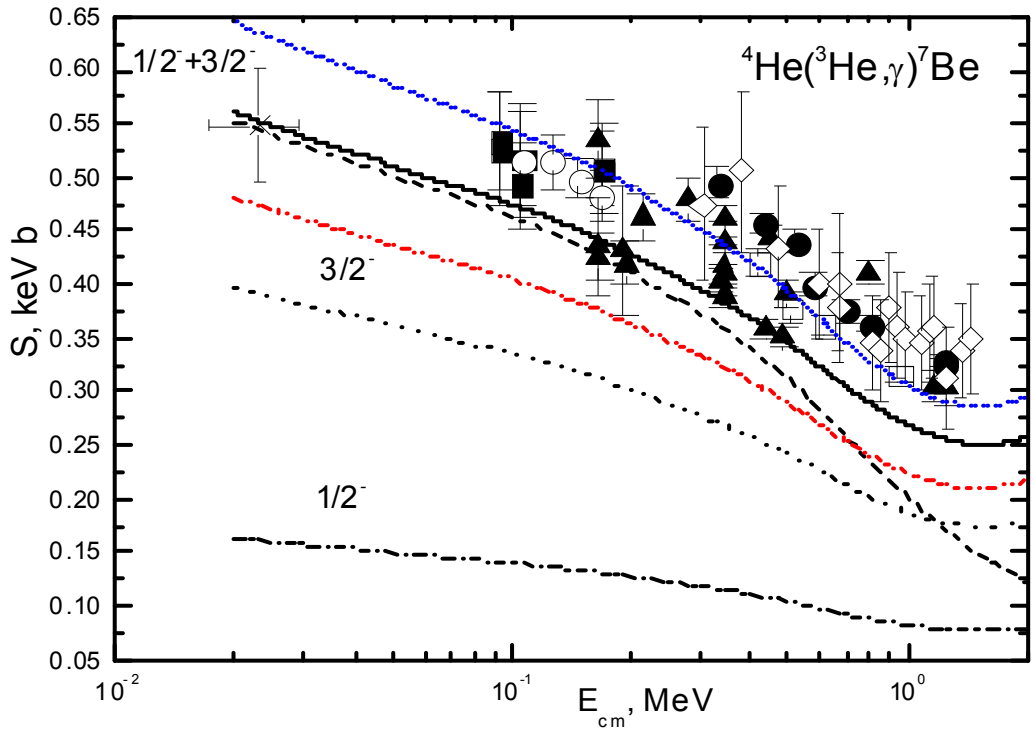

Fig. 13a. Astrophysical $S$-factor of the radiative $^4$He($^3$He, $\gamma$)$^7$Be capture. Filled circles show experimental data Ref. 154, filled squares Ref. 155, open circles Ref. 156, open squares Ref. 157, and filled triangles Ref. 158, open rhombs Ref. 159, crosses Ref. 160. Lines show results of calculation with parameters of the GS and FES potentials from Table 7.

If only transitions from $S$ scattering waves to the ground and first excited bound states are considered, then at 20 keV we obtain value 0.552 keV b (earlier in Refs. 41,136,137 value 0.587 keV b was obtained), which slightly differs from the above value of 0.560 keV b. The calculated $S$-factor for this case is shown in Fig. 13a by the dashed line, which at low energies slightly differs from the previous results obtained taking into account transitions from the $D$ waves. Note that our original calculations of the $^3$H$^4$He $S$-factor leaded to its value of 0.47 keV b.[73,74,123] The difference of this value from the new results[Ошибка! Закладка не определена.,136,137] can also be connected with the imperfection of the ground state potential used in works.[73,74,123]

Using the second variants of the GS potentials and $S$-scattering elastic waves from Table 7 and Table 6, given in brackets, then results for the $S$-factor are shown in Fig. 13a by the red dot-dot-dashed line. The summarized results are shown by the blue dotted line, which lies notably above the available experimental data at lowest energies 23 keV. However, these results better agree with available experimental data lying in their error band at energies from 90 keV to 1.5 MeV. It can be seen from here that just the results for the first variant of the GS potential from Table 7, which slightly understate the AC value, have the better agreement with new experimental data from Ref. 160.

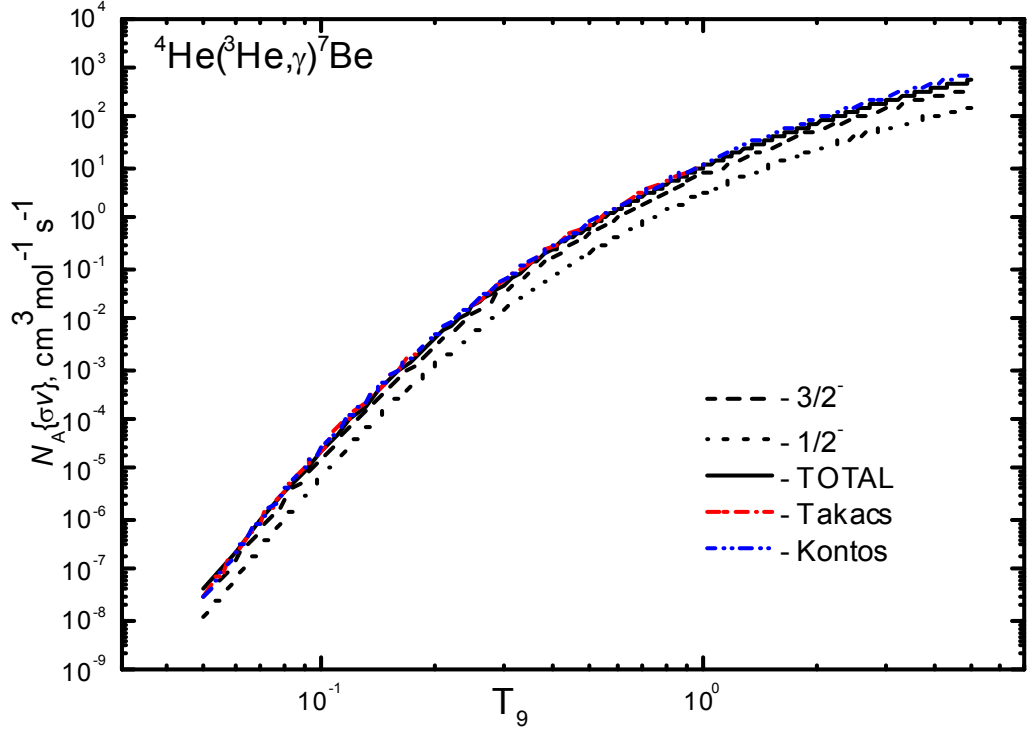

Fig. 13b. Reaction rate of the $^4$He($^3$He, γ)$^7$Be capture at the energy range from 0.05 to 5 $T_9$.

Furthermore, the reaction rates are shown in Fig. 13b at the energy range 0.05 to 5 $T_9$. The capture reaction rate to the FES was shown by the dotted line, to the GS by the dashed line, and the total reaction rate by the solid line, which corresponds to the solid line in Fig. 13a. The reaction rates from Ref. 160 – red dot-dashed line and from Ref. 159 – blue dot-dot-dashed are shown for comparison. The total reaction rate in Fig. 13b can be approximate by the form

$$N_A \langle \sigma v \rangle = 36807.346 / T_9^{2/3} \cdot \exp(-11.354 / T_9^{1/3}) \cdot (1.0 - 15.748 \cdot T_9^{1/3} +$$
$$56.148 \cdot T_9^{2/3} + 27.650 \cdot T_9 - 66.643 \cdot T_9^{4/3} + 21.709 \cdot T_9^{5/3}) +$$
$$44350.648 / T_9^{3/2} \cdot \exp(-16.383 / T_9)$$

which leads to $\chi^2 = 0.9$ at 1% errors of theoretical curve.

### 5.3.3. S-factor of the $^4$He($^2$H, γ)$^6$Li capture

The $E1$ transitions from the $^3P$-scattering waves to the $^3S_1$-ground state of $^6$Li and $E2$ transitions from the $^3D$-scattering waves to the GS are possible for the radiative capture in $^2$He$^4$He channel of $^6$Li. The main contribution to the $E2$ transition at low energies gives the $^3D_3$-wave, having the resonance at 0.71 MeV, and the $E1$ transition is strongly suppressed due to the cluster factor.[74,137] Therefore, the cross section of the radiative capture process, generally, depends on the $E2$ transition. But, as it was shown earlier,[74] at the lowest energies, about 50–150 keV, the contribution of the $E$1 transition becomes dominant.

The results of the $S$-factor calculation for the $^4$He($^2$He, γ)$^6$Li capture more than 5 keV with first variant of the GS potential are shown in Fig. 14a by the solid line. The contribution of the $E$1 transition, which conditioned by the difference of particle masses from integer numbers, is shown by the dashed line. Experimental data is taken from Refs. 163–168. As it is seen from Fig. 14a, the calculation results are in better agreement with data[163,167] at lowest energies. In the range of 5–10 keV for the total calculated $S$-factor determined by the $E$1 and $E$2 transitions, the value of 1.67(1) eV mb was obtained. The contribution of the $E$1 transition in this energy region is defining and comes to 1.39(1) eV mb.[136,137] The mentioned errors of the $S$-factor are determined by averaging over this energy interval. Note that our previous calculations of the $S$-factor of the radiative $^4$He($^2$He, γ)$^6$Li capture led to the value 1.5 eV mb.[73,74,123,126] This difference from the obtained here value 1.67(1) eV mb is also caused by the approximate character of the used earlier potential of the bound state, which was able to obtain the binding energy only with an accuracy only down to few keV.[74]

The presented above results for $S$-factor were obtained by us in Ref. 136 and later get into Refs. 41,137. They were obtained in assumption that contributions of all transitions for $E$1 process can be calculated only with potential of the $^3P_2$ scattering state, which gives the maximal influence to the $S$-factor value. Now we correctly take into account transitions from all $^3P$ scattering states and for the $S$-factor at 5–10 keV obtaining the value 1.64(1) eV·mb, which differs from the previous less than 2%. The contribution of the $E$1 process in the range from 5 to 20 keV now equals 1.37(2) eV·mb. Because in figures it is impossible reflect such changes of the $S$-factor, in Fig. 14a the results for the last variant of calculation with correct accounting of all $^3P$ waves were shown.

Now let us consider the second variant of the GS potential given in Table 7 in brackets, which leads to the dimensionless AC equals 2.93(1). The calculation results of the total $S$-factor are shown in Fig. 14a by the red dotted line and for the $E$1 transition by the blue dot-dashed line. At the lowest energies 5–10 keV, the value of the total $S$-factor equals 1.36(1) eV·mb, and for the $E$1 transition equals 1.12(1) eV·mb.

Note that the results of extrapolation of experimental data given in, for example, work Ref. 66 at 10 keV yield 1.6(1) eV mb, which well agrees with the value that we obtained first. However, the following value was obtained in theoretical calculations of Ref. 141 for the $S$-factor at zero energy: 1.2(1) eV mb, which corresponds to an asymptotic constant of 3.06, and the constant extracted from experimental data[140,167] is equal to 2.96(14). The second variant of the GS potential, shown in Table 7 in brackets, is better corresponded to this value. The value 1.60(17) eV mb was obtained in the latest work for extracting $S$-factor at the zero energy from the experimental data,[169] which in a good agreement with our first result.

The reaction rates are shown in Fig. 14b at the energy range 0.05 to 5 $T_9$. The solid line shows the reaction rate with the first variant of the GS potential from Table 7. The dashed line shows the rate with the second variant of the potential from Table 7 with depth of 92 MeV. Besides, the results of Ref. 170 are shown by the red dot-dashed line and the result of analysis[107] by the blue dotted line. The total reaction rate for the potential of the depth 75 MeV shown in the Fig. 14b can be approximated by the form

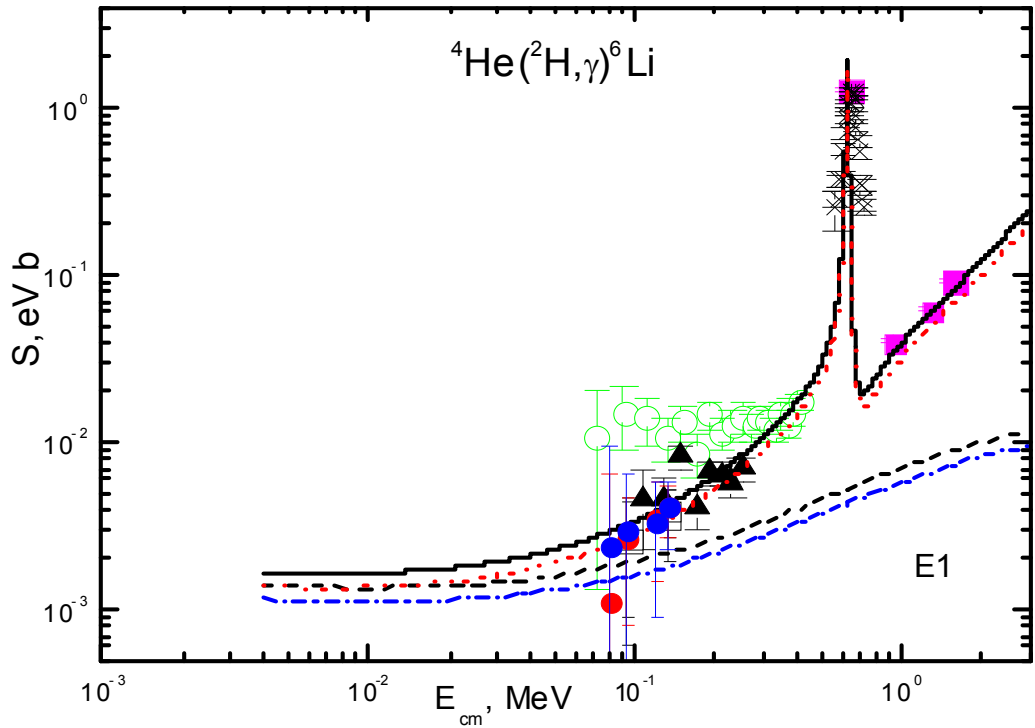

Fig. 14a. Astrophysical *S*-factor of the radiative $^4$He($^2$He, $\gamma$)$^6$Li capture. Filled squares show experimental data Ref. 163, crosses Ref. 165, open circles Ref. 166, triangles – Refs. 163,167, open squares Ref. 168, and points – Ref. 153. The dashed line shows the contribution of *E*1 process for transition to the ground state from the *P* scattering waves. The solid line shows the total *S*(*E*1+*E*2)-factor.

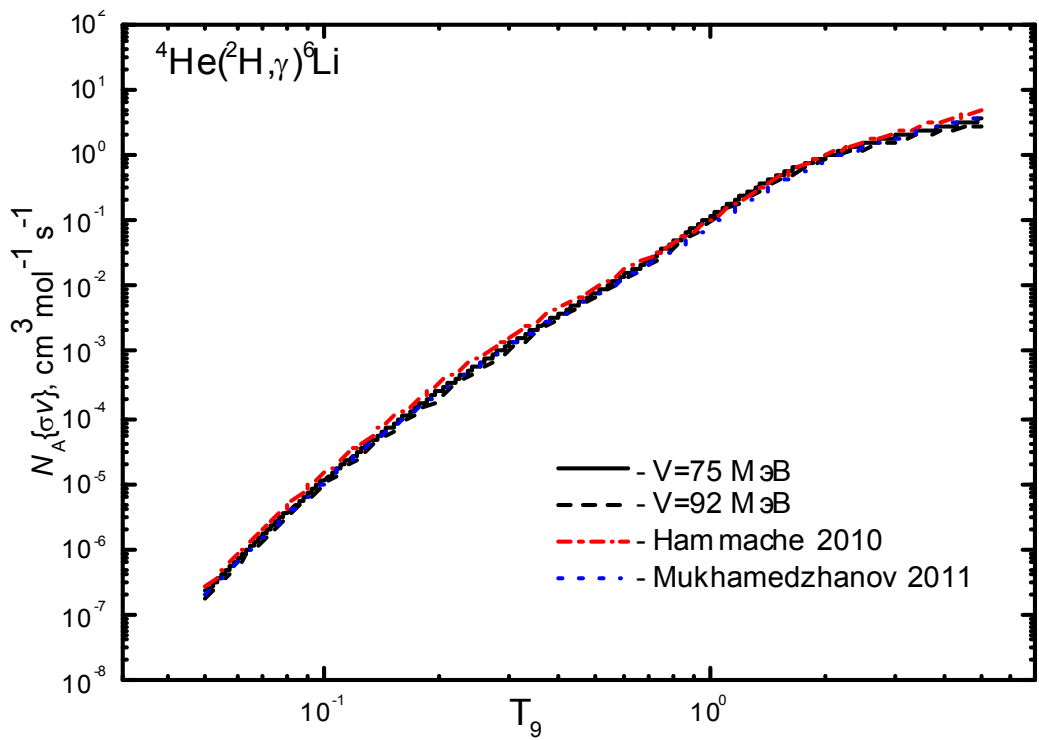

Fig. 14b. Reaction rate of the $^4$He($^2$He, $\gamma$)$^6$Li capture at the energy range from 0.05 to 5 $T_9$.

$$N_A \langle \sigma v \rangle = 17.128/T_9^{2/3} \cdot \exp(-7.266/T_9^{1/3}) \cdot (1.0 - 4.686 \cdot T_9^{1/3} +$$
$$15.877 \cdot T_9^{2/3} - 21.523 \cdot T_9 + 18.703 \cdot T_9^{4/3} - 4.554 \cdot T_9^{5/3}) + 53.817/T_9^{3/2} \cdot \exp(-6.933/T_9)$$

which leads to $\chi^2 = 0.08$ at 1% errors of theoretical curve. The total reaction rate for the potential of the depth 92 MeV shown in the Fig. 14b can be approximated by the form

$$N_A \langle \sigma v \rangle = 13.206/T_9^{2/3} \cdot \exp(-7.272/T_9^{1/3}) \cdot (1.0 - 4/381 \cdot T_9^{1/3} +$$
$$15.179 \cdot T_9^{2/3} - 20.728 \cdot T_9 + 18.800 \cdot T_9^{4/3} - 4.658 \cdot T_9^{5/3}) + 49.399/T_9^{3/2} \cdot \exp(-6.950/T_9)$$

with the same $\chi^2$.

## 6. Conclusions

### 6.1. $^3H(p,\gamma)^4He$ reaction

#### 6.1.1. Nuclear physics

Thereby, in the framework of considered modified potential cluster model based on the intercluster potentials describing elastic scattering phase shifts and characteristics of the binding state with the potential parameters suggested about 20 years ago,[101] on the basis of only the $E1$ transition we succeeded in description of the general behavior of the $S$-factor of the proton capture on $^3H$ at energies from 50 to 700 keV. Really, on the basis of analysis of the experimental data above 700 keV[102] about 20 years ago we have done calculations of the $S$-factor for energies down to 10 keV.[101] As we can see it now, the results of these calculations reproduce well new data on the $S$-factor, obtained in Ref. 9 (points in Fig. 2) at energies in the range 50 keV to 5 MeV.

However the available experimental data on the $S$-factor at 50 keV and below have a low accuracy and significant ambiguity, as it seen from Fig. 2. To avoid these ambiguities, there is a need for new additional and independent measurements of the $S$-factor in the energy range from about 5–10 to 30–50 keV with minimal errors. No new experimental data in this energy range has been forthcoming for more than 10 years.[10] Reliable measurements of $S$-factor at energies of 50 keV–5.0 MeV were made more than 20 years ago.[9] Evidently, modern measurement techniques could reduce error values and obtain more reliable data, especially at the lowest energies. This, in turn, will eliminate the existing ambiguities in determining the reaction rate.

#### 6.1.2 Nuclear astrophysics

The magnitude of the $^3H(p, \gamma)^4He$ capture reaction rate calculated in this paper at temperatures from 0.01 $T_9$ up to 5 $T_9$ leads to the conclusion that this reaction will not make some contribution to the formation of $^4He$ nuclei in the primordial nucleosynthesis of elements in the Universe. But the results obtained for the reaction rate using a simple numerical approximation could be of use in determining the yield of $^4He$ in this reaction and thus help to estimate the abundance of helium nuclei formed in the primordial nucleosynthesis of the Universe.

At the same time the available errors of measurements of the astrophysical $S$-

factor[10] may significantly affect the value of the reaction rate of the radiative proton capture on $^3$H leading to ambiguities in calculations of $^4$He yield and, ultimately, affect the results obtained for its abundance.

## 6.2. $^3He(^2H,\gamma)^5Li$ reaction

These comparatively simple model representations succeeded in obtaining theoretical results in general agreement with the available experimental data for the $S$-factor, except for the first three points at 185–235 keV and at energies higher than 600–800 keV.

The minimum value 14.5(2.8) $\mu b$ of the experimental cross section is close to 1 MeV,[31] and the calculated value is 11.5 $\mu b$. We can identify no reasonable explanation for this discrepancy between theory and experiment as yet. Although the 12 transitions listed in Table 5 were taken into account, we did not consider any other features of this process.

For example, the small admixture of the $^4F_{3/2}$ component compared with the dominant $^4P_{3/2}$ in GS of $^5$Li may have an input into the total cross section. We postulate that this may increase the total cross-sectional value, particularly at higher energies.

Another option for improving the agreement of the obtained results with experiment is the inclusion of a small admixture of the $\alpha N$ component in the $d\tau$ cluster channel wave function of $^5$Li, which may affect the asymptotics of the radial function, and as a consequence should redistribute the space probability density in the nuclear interior. The same superposition of various cluster components is used in RGM.[64,65]

The result may be important in practical terms with regard to the obtained approximation of the experimental $S$-factor with energy constant 0.39 keV b and mean $\chi^2 = 0.21$ at 19% experimental error bars below 300 keV. Another approximation using the square form (27) was found at energies below 250 keV; this gives $\chi^2 = 0.17$ with 1% calculation errors. Within an energy range of between 185 keV and 1.4 MeV, a fit using a Breit-Wigner resonance-type formula led to $\chi^2 = 0.11$ with 19% experimental error bars. The same formula applied to the approximation of the theoretical curve in the energy interval 20 keV to 1.5 MeV gave $\chi^2 = 14.5$ with a 1% calculation error.

For the reaction rate $\sigma v$ we suggest parametrization (28) for temperatures between 0.03 and 3 $T_9$, and would recommend this for the solution of several applied problems in current astrophysical and fusion research.

Finally, it should be noted that we are aware of only one measurement of the total cross section of this capture reaction, as presented in Ref. 31; this was performed in the late 1960s, that is, about 50 years ago, and may be subject to change. A firm conclusion would therefore require more precise measurements of this reaction, using modern methods and obtained irrespective of previous data. We hope that the results presented here may act as a guide for future experimental proposals.

## 6.3. $^4He(^3He, \gamma)^7Be$, $^4He(^3H, \gamma)^7Li$ and $^4He(^2H, \gamma)^6Li$ capture reactions

### 6.3.1. Nuclear physics

Thus, specified calculation variants of the $S$-factor, when the required potentials are used for all partial waves, better agree with the available earlier and new experimental data.

Results for the $^4\text{He}(^2\text{H}, \gamma)^6\text{Li}$ capture with the second GS potential, evidently, better coincide with new data.[168] However, the experimental errors are so high that it is impossible to unambiguously prefer one of the GS potentials. As concerns the $^4\text{He}(^3\text{He}, \gamma)^7\text{Be}$ capture, so the calculation results that was done by us in 2010 Ref. 136 and specified by 6% in this work, agree within limits of errors with new measurements from Ref. 160 at energy 23 keV.

### *6.3.2 Nuclear astrophysics*

The slightly refined variants of calculations for the astrophysical *S*-factor of the $^3\text{He}(^4\text{He},\gamma)^7\text{Be}$ reaction are in better agreement with the data available earlier and the latest experimental data. The new lowest measured point $S(23^{+6}_{-5}\,\text{keV})=0.548\pm0,054\,\text{keV}\cdot\text{b}$ (Takács *et al.*, 2015 Ref. 159) is just on the theoretical curve calculated earlier by us for *S*(E). So, the predictive reliability of the developing cluster model approach was demonstrated. The new parametrization for the reaction rate is obtained and may be recommended for the astrophysical evaluation of the $^7\text{Be}$ production. Also we ought to note that at the beginning of this year the astrophysical *S*-factor of the $^3\text{He}(^4\text{He},\gamma)^7\text{Be}$ reaction was published in Ref. 171, but the reaction rate and its parametrization were not calculated.

We would like to note that a preliminary version of the present results Ref. 172 has already found successful application; in particular, the reaction rate analytical parametrization was used in the latest calculations on the solution of the primordial lithium abundance problem of BBN Ref. 173.

### Acknowledgments

This work was supported by the Ministry of Education and Science of the Republic of Kazakhstan (Grant No. AP05130104) entitled "Study of Reactions of the Radiative Capture in Stars and in the Controlled Thermonuclear Fusion" through the Fesenkov Astrophysical Institute of the National Center for Space Research and Technology of the Ministry of Defence and Aerospace Industry of the Republic of Kazakhstan (RK).

### References

1. S. B. Dubovichenko and A. V. Dzhazairov-Kakhramanov, *Int. Jour. Mod. Phys. E* **21**, 1250039(1-44) (2012).
2. S. B. Dubovichenko, A. V. Dzhazairov-Kakhramanov and N. V. Afanasyeva, *Int. Jour. Mod. Phys. E* **22**, 1350075(1-53) (2013).
3. S. B. Dubovichenko, A. V. Dzhazairov-Kakhramanov and N. A. Burkova, *Int. Jour. Mod. Phys. E* **22**, 1350028(1-52) (2013).
4. S. B. Dubovichenko and A. V. Dzhazairov-Kakhramanov, *Int. Jour. Mod. Phys. E* **23**, 1430012(1-55) (2014).
5. S. B. Dubovichenko and A. V. Dzhazairov-Kakhramanov, *Int. Jour. Mod. Phys. E* **26**, 1630009(1-56) (2017).
6. C. A. Barnes, D. D. Clayton, D. N. Schramm, *Essays in Nuclear Astrophysics. Presented to William A. Fowler* (Cambridge University Press, Cambridge, 1982).
7. D. S. Gorbunov, V. A. Rubakov, *Introduction to theory of early Universe. Hot Big Bang theory* (World Scientific, Singapore, 2011).

8. Ya. B. Zeldovich, I. D. Novikov, *The Structure and Evolution of the Universe* (University of Chicago Press, Chicago, 1983).
9. K. Hahn et al. Phys. Rev. C **51** 1624 (1995).
10. R. Canon *et al.*, *Phys. Rev. C* **65** 044008.1 (2002).
11. T. Shima *et al.*, *Phys. Rev. C* **72** 044004 (2005); B. Nilsson *et al.*, *Phys. Rev. C* **75** 014007 (2007).
12. D. Halderson, *Phys. Rev. C* **70** 034607 (2004); Nir Nevo Dinur, Nir Barnea, Winfried Leidemann, *Few-Body Systems* **55** 997 (2014); K. M. Nollett, S. Burles, *Phys. Rev. D* **61** 123505 (2000).
13. W. Horiuchi, Y. Suzuki, and K. Arai, *Phys. Rev. C* **85** 054002 (2012); S. Quaglioni *et al.*, *Phys. Rev. C* **69** 044002 (2004).
14. V. A. Bednyakov, *Phys. Part. Nucl.* **33** 915 (2002).
15. Ya. M. Kramarovskii, V. P. Chechev, *Phys. Usp.* **42** 563 (1999); http://iopscience.iop.org/article/10.1070/PU1999v042n06ABEH000559/meta
16. S. B. Dubovichenko, A. V. Dzhazairov-Kakhramanov, N. V. Afanasyeva, *Nucl. Phys. A* **963**, 52 (2017).
17. Yu. E. Penionzhkevich, *Phys. Part. Nucl.* **43** 452 (2012).
18. T. S. Suzuki *et al.*, *Astroph. Journ. Lett.* **439** L59 (1995).
19. C. Casella *et al.*, *Nucl. Phys. A* **706** 203 (2002).
20. S. B. Dubovichenko, A. V. Dzhazairov-Kakhramanov, *Nucl. Phys. A* **941** 335 (2015).
21. D. R. Tilley, H. R. Weller, G. M. Hale, *Nucl. Phys. A* **541** 1 (1992).
22. C. R. Brune *et al.*, *Phys. Rev. C* **60** 015801 (1999).
23. S. B. Borzakov *et al.*, *Sov. J. Nucl. Phys.* **35** 307 (1982).
24. R. Wervelman *et al.*, *Nucl. Sci. & Eng. (NSE)* **102** 428 (1989).
25. A. Krauss *et al.*, *Nucl. Phys. A* **465** 150 (1987).
26. S. L. Greene, *UCRL-70522* (1967).
27. R. E. Brown, N. Jarmie, *Phys. Rev. C* **41** 1391 (1990).
28. M. Aliotta *et al.*, *Nucl. Phys. A* **690** 790 (2001).
29. L. Stewart, G. M. Hal, *LA-5828-MS* (1975).
30. Zhou Jing *et al.*, *Chin. Phys. C* **33** 350 (2009).
31. W. Buss *et al.*, *Nucl. Phys. A* **112** 47 (1968).
32. Table of Isotopic Masses and Natural Abundances, (1999); https://www.ncsu.edu/ncsu/pams/chem/msf/pdf/IsotopicMass_NaturalAbundance.pdf
33. J. E. Purcell *et al.*, *Nucl. Phys. A* **848** 1 (2010).
34. J. M. Eisenberg, W. Greiner, *Excitation mechanisms of the nucleus* (North Holland, Holland, 1976).
35. G. R. Caughlan, W. A. Fowler, *Atom. Dat. & Nucl. Dat. Tab.* **40** 283 (1988).
36. P. D. Serpico *et al.*, *Journal of Cosmology and Astroparticle Physics (JCAP)* **12** 010 (2004).
37. C. A. Bertulani, T. Kajino, *Prog. Part. Nucl. Phys.* **89** 56 (2016).
38. W. A. Fowler, G. R. Caughlan, B. A. Zimmerman, *Annu. Rev. Astron. Astrophys (ARA&A)* **13** 69 (1975).
39. M. J. Harris, W. A. Fowler, G. R. Caughlan, B. A. Zimmerman, *Annu. Rev. Astron. Astrophys (ARA&A)* **21** 165 (1983).
40. G. R. Caughlan, W. A. Fowler, M. J. Harris, B. A. Zimmerman, *Atom. Dat. & Nucl. Dat. Tab.* **32** 197 (1985).
41. S. B. Dubovichenko, *Thermonuclear Processes in Stars and Universe* (Lambert,

Germany, 2015); https://www.scholars-press.com/catalog/details/store/gb/book/978-3-639-76478-9/Thermonuclear-processes-in-stars
42. S. B. Dubovichenko, A. V. Dzhazairov-Kakhramanov, *Euro Phys. J. A* **39** 139 (2009).
43. S. B. Dubovichenko, A. V. Dzhazairov-Kakhramanov, *Annal. der Phys.* **524** 850 (2012).
44. S. B. Dubovichenko, A. V. Dzhazairov-Kakhramanov, *Astrophys. J.* **819** 78 (8pp) (2016); doi: 10.3847/0004-637X/819/1/78.
45. S. B. Dubovichenko, N. Burtebaev, A. V. Dzhazairov-Kakhramanov, D. Alimov, *Mod. Phys. Lett. A* **29** 1450125(1-16) (2014).
46. G. Imbriani, In: Proceedings of Third Europ. Summ. Sch. on Exp. Nucl. Astrop., S. Tecla (Catania), Sicily, Italy, October 2-9 (2005).
47. A. Caciolli *et al.*, *Eur. Phys. J. A* **39** 179 (2009).
48. K. Wildermuth and Y. C. Tang, *A unified theory of the nucleus* (Vieweg, Branschweig, 1977).
49. T. Mertelmeir and H. M. Hofmann, *Nucl. Phys. A* **459** 387 (1986).
50. J. Dohet-Eraly, *Microscopic cluster model of elastic scattering and bremsstrahlung of light nuclei* (Université Libre De Bruxelles, Bruxelles, 2013); http://theses.ulb.ac.be/ETD-db/collection/available/ULBetd-09122013-100019/unrestricted/these_Jeremy_Dohet-Eraly.pdf.
51. J. Dohet-Eraly and D. Baye, *Phys. Rev. C* **84** 014604 (2011).
52. P. Descouvemont and M. Dufour, *Microscopic cluster model*, in: Clusters in Nuclei, Second edition by C. Beck, (Springer-Verlag, Berlin Heidelberg, 2012).
53. P. Descouvemont, *Microscopic cluster models. I.*; available online at: http://www.nucleartheory.net/Talent_6_Course/TALENT_lectures/pd_microscopic_1.pdf
54. A. V. Nesterov, F. Arickx, J. Broeckhove and V. S. Vasilevsky, *Phys. Part. Nucl.* **41** 716 (2010).
55. A. V. Nesterov, V. S. Vasilevsky and T. P. Kovalenko, *Ukr. J. Phys.* **58** 628 (2013).
56. O. F. Nemets, V. G. Neudatchin, A. T. Rudchik, Y. F. Smirnov and Yu. M. Tchuvil'sky, *Nucleon Association in Atomic Nuclei and the Nuclear Reactions of the Many Nucleons Transfers* (Naukova dumka, Kiev, 1988).
57. S. B. Dubovichenko, *Radiative neutron capture and primordial nucleosynthesis of the Universe.* 5th Edition, corrected and added (Lambert Academy Publ. GmbH&Co. KG., Saarbrucken, 2016) 496p.; https://www.ljubljuknigi.ru/store/ru/book/radiative_neutron_capture/isbn/978-3-659-82490-6 (in Russian)
58. S. B. Dubovichenko, *Thermonuclear Processes of the Universe* (NOVA Sci. Publ., New-York, 2012) 196p.; https://www.novapublishers.com/catalog/product_info.php?products_id=31125
59. S. E. Sharapov, T. Hellsten, and V. G. Kiptily, *Nucl. Fusion* **56** 112021(10pp) (2016).
60. Experimental Nuclear Reaction Data (EXFOR), Database Version of 2017-05-10, https://www-nds.iaea.org/exfor/exfor.htm.
61. CODATA 2014, Fundamental Physical Constants, http://physics.nist.gov/cgi-bin/cuu/Category?view=html&Atomic+and+nuclear.x=100&Atomic+and+nuclear.y=13
62. Centre for photonuclear experiments data, http://cdfe.sinp.msu.ru/

63. S. B. Dubovichenko, and N. A. Burkova, *Mod. Phys. Lett. A* **29** 1450036(1-14) (2014).
64. P. Descouvemont *et al*., *AIP Advances* **4** 041011 (2014).
65. P. Descouvemont *et al*., *The Universe Evolution. Astrophysical and Nuclear Aspects* (NOVA Sci. Publ., New-York, 2013) P.1-48.
66. C. Angulo *et al.*, *Nucl. Phys. A* **656** 3 (1999).
67. S. B. Dubovichenko and A. V. Dzhazairov-Kakhramanov, *The Universe Evolution. Astrophysical and Nuclear Aspects* (NOVA Sci. Publ., New-York, 2013) P.49-108.
68. G. R. Plattner and R. D. Viollier, *Nucl Phys A* **365** 8 (1981).
69. R. Neugart *et al.*, *Phys. Rev. Lett.* **101** 132502 (2008).
70. T. Tombrello and P. D. Parker, *Phys. Rev*. **131** 2578 (1963).
71. T. Tombrello and P. D. Parker, *Phys. Rev*. **131** 2582 (1963).
72. V. Guimarães *et al.*, International Symposium on Nuclear Astrophysics, Nuclei in the Cosmos IX (CERN, Geneva, Switzerland, June 25-30, 2006) P.108.
73. S. B. Dubovichenko and A. V. Dzhazairov-Kakhramanov, *Phys. Atom. Nucl.* **58** 579 (1995).
74. S. B. Dubovichenko and A. V. Dzhazairov-Kakhramanov, *Phys. Part. Nucl.* **28** 615 (1997).
75. D. A. Varshalovich, A. N. Moskalev and V. K. Khersonskii, *Quantum Theory of Angular Momentum* (World Scientific, Singapore, 1989).
76. Fundamental Physical Constants (2010), (NIST Ref. on Constants, Units and Uncertainty), proton magnetic moment, http://physics.nist.gov/cgi-bin/cuu/Value?mup#mid
77. M. P. Avotina and A. V. Zolotavin, *Moments of the ground and excited states of nuclei* (Atomizdat, Moscow, 1979).
78. S. B. Dubovichenko, *Phys. Part. Nucl*. **44** 803 (2013).
79. Fundamental Physical Constants (2010), (NIST Ref. on Constants, Units and Uncertainty), proton mass, http://physics.nist.gov/cgi-bin/cuu/Value?mpu#mid
80. Nuclear Wallet Cards database ($^2$H isotope), (2012); http://cdfe.sinp.msu.ru/cgi-bin/gsearch_ru.cgi?z=1&a=2
81. Nuclear Wallet Cards database ($^3$H isotope), (2012); http://cdfe.sinp.msu.ru/cgi-bin/gsearch_ru.cgi?z=1&a=3
82. Nuclear Wallet Cards database ($^3$He isotope), (2012); http://cdfe.sinp.msu.ru/cgi-bin/gsearch_ru.cgi?z=2&a=3
83. Nuclear Wallet Cards database ($^4$He isotope), (2012); http://cdfe.sinp.msu.ru/cgi-bin/gsearch_ru.cgi?z=2&a=4
84. Nuclear Wallet Cards database ($^6$Li isotope), (2012); http://cdfe.sinp.msu.ru/cgi-bin/gsearch_ru.cgi?z=3&a=6
85. Nuclear Wallet Cards database ($^7$Li isotope), (2012); http://cdfe.sinp.msu.ru/cgi-bin/gsearch_ru.cgi?z=3&a=7
86. Nuclear Wallet Cards database ($^7$Be isotope), (2012); http://cdfe.sinp.msu.ru/cgi-bin/gsearch_ru.cgi?z=4&a=7
87. V. G. Neudatchin *et al*., *Phys. Rev. C* **45** 1512 (1992).
88. V. G. Neudatchin, A. A. Sakharuk, S. B. Dubovichenko, *Few-Body Sys*. **18** 159 (1995).
89. C. Itzykson, M. Nauenberg, *Rev. Mod. Phys*. **38** 95 (1966).
90. A. Bohr, B. R. Mottelson, *Nuclear structure Vol. I. Single-particle motion* (World Scientific Publ. Co. Ldt., Singapore, 1998).

91. S. B. Dubovichenko, *Phys. At. Nucl.* **74** 358 (2011).
92. T. A. Tombrello, *Phys. Rev.* **138** B40 (1965).
93. Y. Yoshino *et al.*, *Prog. Theor. Phys.* **103** 107 (2000).
94. D. H. McSherry, S. D. Baker, *Phys. Rev. C* **1** 888 (1970).
95. L. Drigo, G. Pisent, *Nuovo Cim. B* **LI** 419 (1967).
96. G. Szaloky, F. Seiler, *Nucl. Phys. A* **303** 57 (1978).
97. T. A. Tombrello *et al.*, *Nucl. Phys.* **39** 541 (1962).
98. J. S. McIntosh, R. L. Gluckstern, S. Sack, *Phys. Rev.* **88** 752 (1952).
99. R. M. Frank, J. L. Gammel, *Phys. Rev.* **99** 1406 (1955).
100. R. Kankowsky *et al.*, *Nucl. Phys. A* **263** 29 (1976).
101. S. B. Dubovichenko, Phys. At. Nucl. **58** 1295 (1995).
102. Yu. M. Arkatov *et al.*, *Sov. J. Nucl. Phys.* **12** 227 (1970).
103. E. G. Adelberger *et al.*, *Rev. Mod. Phys.* **83** 195 (2011).
104. CODATA 2014, Fundamental Physical Constants, http://physics.nist.gov/cgi-bin/cuu/Category?view=html&Atomic+and+nuclear.x=87&Atomic+and+nuclear.y=12
105. T. K. Lim, *Phys. Lett. B* **55** 252 (1975); T. K. Lim, *Phys. Lett. B* **44** 341 (1973).
106. N. K. Timofeyuk, *Phys. Rev. C* **81** 064306(1-21) (2010).
107. A. M. Mukhamedzhanov, L. D. Blokhintsev and B. F. Irgaziev, *Phys. Rev. C* **83** 055805(1-9) (2011).
108. B. F. Gibson, *Nucl. Phys. A* **353** 85 (1981).
109. J. E. Perry, S. J. Bame, *Phys. Rev.* **99** 1368 (1955).
110. F. Balestra *et al.*, *Nuovo Cim.* **38A** 145 (1977).
111. W. E. Meyerhof *et al.*, *Nucl. Phys. A* **148** 211 (1970).
112. G. Feldman *et al.*, *Phys. Rev. C* **42** R1167 (1990).
113. P. E. Hodgson, *The Optical model of elastic scattering* (Clarendon Press, Oxford, 1963).
114. S. B. Dubovichenko, N. A. Burkova, A. V. Dzhazairov-Kakhramanov and A. S. Tkachenko, *Rus. Phys. J.* **60** 935 (2017).
115. B. Jenny *et al.*, *Nucl. Phys. A* **337** 77 (1980).
116. S. B. Dubovichenko and A. V. Dzhazairov-Kakhramanov, *Sov. J. Nucl. Phys.* **51** 971 (1990).
117. S. B. Dubovichenko, *Phys. Atom. Nucl.* **58** 1174 (1995).
118. D. R. Tilley *et al.*, *Nucl. Phys. A* **708** 3 (2002).
119. V. I. Kukulin, V. G. Neudatchin and Yu. F. Smirnov, *Fiz. Elem. Chastits At. Yadra* **10** 1236 (1979).
120. V. G. Neudatchin, A. A. Sakharuk and Yu. F. Smirnov, *Fiz. Elem. Chastits At. Yadra* **23** 480 (1992).
121. V. G. Neudatchin, B. G. Struzhko and V. M. Lebedev, *Fiz. Elem. Chastits At. Yadra* **36** 888 (2005).
122. S. B. Dubovichenko, V. G. Neudatchin, A. A. Sakharuk, and Yu. F. Smirnov, *Bull. Acad. Sci. USSR Ser. Fiz.* **54** 911 (1990).
123. S. B. Dubovichenko, A. V. Dzhazairov-Kakhramanov, *Phys. Atom. Nucl.* **58** 788 (1995).
124. S. B. Dubovichenko, A. V. Dzhazairov-Kakhramanov, *Yad. Fiz.* **56** 87 (1993).
125. S. B. Dubovichenko, A. V. Dzhazairov-Kakhramanov, *Phys. Atom. Nucl.* **57** 733 (1994).
126. S. B. Dubovichenko, *Caracteristics of light atomic nuclei in the potential cluster model* (Daneker, Almaty, 2004); arXiv:1006.4944 [nucl-th].

127. S. B. Dubovichenko, N. A. Burkova, A. V. Dzhazairov-Kakhramanov, *Indian J Phys*. (2018); doi: 10.1007/s12648-018-1287-0.
128. A. C. Barnard, C. M. Jones, G. C. Phillips, *Nucl. Phys*. **50** 629 (1964).
129. R. Spiger, T. A. Tombrello, *Phys. Rev*. **163** 964 (1967).
130. M. Ivanovich, P. G. Young, G. G. Ohlsen, *Nucl. Phys. A* **110** 441 (1968).
131. L. C. McIntyre, W. Haeberli, *Nucl. Phys. A* **91** 382 (1967).
132. L. G. Keller, W. Haeberli, *Nucl. Phys. A* **156** 465 (1979).
133. W. Gruebler *et al.*, *Nucl. Phys. A* **242** 265 (1975).
134. P. A. Schmelzbach *et al.*, *Nucl. Phys. A* **184** 193 (1972).
135. S. B. Dubovichenko, *Phys. Atom. Nucl*. **61** 162 (1998).
136. S. B. Dubovichenko, *Phys. Atom. Nucl*. **73**, 1526 (2010).
137. S. B. Dubovichenko and Yu. N. Uzikov, *Phys. Part. Nucl*. **42** 251 (2011).
138. S. B. Dubovichenko, *Calculation methods of nuclear characteristics* (Complex, Almaty, 2006); arXiv:1006.4947 [nucl-th].
139. Chart of nucleus shape and size parameters ($0 \leq Z \leq 14$), http://cdfe.sinp.msu.ru/cgi-bin/muh/radchartnucl.cgi?zmin=0&zmax=14&tdata=123456&selz=3&sela=9
140. L. D. Blokhintsev *et al.*, *Phys. Rev. C* **48** 2390 (1993).
141. L. D. Blokhintsev *et al.*, *Yad. Fiz*. **69** 456 (2006).
142. L. D. Blokhintsev, I. Borbei, E. I. Dolinski, *Fiz. Elem. Chastits At. Yadra* **8** 1189 (1977).
143. S. B. Igamov, R. Yarmukhamedov, *Nucl. Phys. A* **781** 247 (2007).
144. Brune C.R. *et al.*, *Phys. Rev. Lett*. **83** 4025 (1999).
145. S. B. Igamov, K. I. Tursunmakhatov, R. Yarmukhamedov, arXiv:0905.2026v4 [nucl-th] 6 Jan. 2010. 28p.
146. K. Langanke, *Nucl. Phys. A* **457** 351 (1986).
147. T. Kajino, *Nucl. Phys. A* **460** 559 (1986).
148. N. A. Burkova *et al.*, *Phys. Lett. B* **248** 15 (1990).
149. C. R. Brune, R. W. Kavanagh, C. Rolf, *Phys. Rev. C* **50** 2205 (1994).
150. G. M. Griffiths *et al.*, *Can. J. Phys*. **39** 1397 (1961).
151. Y. Tokimoto *et al.*, *Phys. Rev. C* **63** 035801(20) (2001).
152. U. Schroder *et al.*, *Phys. Lett. B* **192** 55 (1987).
153. D. Trezzi *et al.*, Astroparticle Physics **89** 57 (2017).
154. T. A. D. Brown *et al.*, *Phys. Rev. C* **76** 055801 (2007); arXiv:0710.1279v4 [nucl-ex] 5 Nov. 2007.
155. F. Confortola *et al.*, *Phys. Rev. C* **75** 065803 (2007); arXiv:0705.2151v1 [nucl-ex] 15 May 2007.
156. G. Gyurky *et al.*, *Phys. Rev. C* **75** 035805 (2007).
157. N. Singh *et al.*, *Phys. Rev. Lett*. **93** 262503 (2004).
158. J. L. Osborn *et al.*, *Nucl. Phys. A* **419** 15 (1984).
159. A. Kontos *et al.*, *Phys. Rev. C* **87** 065804(9p.) 2013.
160. M. P. Takács *et al.*, *Phys. Rev. D* **91** 123526(7p.) 2015.
161. D. Bemmerer *et al.*, *Phys. Rev. Lett*. **97** 122502 (2006); arXiv:nucl-ex/0609013v1 11 Sep. 2006.
162. H. Costantini *et al.*, *Nucl. Phys. A* **814** 144 (2008).
163. R. Yarmukhamedov *et al.*, *Uzbek. J. Phys*. **12** 233 (2010).
164. R. C. Robertson *et al.*, *Phys. Rev. Lett*. **47** 1867 (1981).
165. P. Mohr *et al.*, *Phys. Rev. C* **50** 1543 (1994).
166. J. Kiener *et al.*, *Phys. Rev. C* **44** 2195 (1991).

167. R. Yarmukhamedov, *Nucl. Phys. A* **673** 509 (2000).
168. M. Anders *et al.*, *Phys. Rev. Lett.* **113** 042501 (2014).
169. S. B. Igamov, *Phys. Atom. Nucl.* **80** 251 (2017).
170. F. Hammache *et al.*, *Phys. Rev. C* **82** 065803 (2010).
171. E M Tursunov, S A Turakulov and A S Kadyrov *Phys. Rev. C* **97** 035802 (2018).
172. S B Dubovichenko, N A Burkova, A V Dzhazairov-Kakhramanov arXiv: 1706.05245v1 [nucl-th].
173. Vinay Singh et al arXiv:1708.05567v2 [nucl-th].